\documentclass[11pt]{article}

\usepackage[utf8]{inputenc}
\usepackage[T1]{fontenc}
\usepackage[margin=1in]{geometry}
\usepackage[authoryear,round]{natbib}
\usepackage{graphicx}
\usepackage{caption}
\usepackage{subcaption}
\usepackage{amsmath,amsfonts,amssymb,mathtools}
\usepackage{float}
\usepackage[version=3]{mhchem}
\usepackage{xcolor}
\usepackage{multirow}
\usepackage{dcolumn}
\usepackage{hyperref}
\usepackage{indentfirst}

\hypersetup{
  colorlinks=true,
  linkcolor=blue,
  citecolor=blue,
  urlcolor=blue
}

\newcommand{\tablefoot}[1]{%
  \par\smallskip
  \begin{minipage}{0.98\linewidth}\footnotesize #1\end{minipage}%
}

\title{\vspace{-2.2cm} \textbf{Contrasting C/O ratios in Uranus and Neptune from disequilibrium chemistry: A clue to distinct evolutionary pathways?}}

\author{\parbox{\textwidth}{
T. Briand$^{1}$, V. Hue$^{1}$, O. Mousis$^{2}$, T. Cavali\'e$^{3,4}$,
T. Benest Couzinou$^{5}$, A. Schneeberger$^{6}$, and M. D. Hofstadter$^{7}$\\[0.9em]
\small $^{1}$Aix-Marseille Universit\'e, CNRS, CNES, Institut Origines, LAM, Marseille, France\\
\small $^{2}$Solar System Science and Exploration Division, Southwest Research Institute, 1301 Walnut St, Ste 400, Boulder, CO, USA\\
\small $^{3}$Laboratoire d'Astrophysique de Bordeaux, Univ. Bordeaux, CNRS, B18N, all\'ee Geoffroy Saint-Hilaire, 33615 Pessac, France\\
\small $^{4}$LIRA, Observatoire de Paris, Universit\'e PSL, CNRS, Sorbonne Universit\'e, Universit\'e Paris Cit\'e, 5 place Jules Janssen, 92195 Meudon, France\\
\small $^{5}$National Key Laboratory of Deep Space Exploration/State Key Laboratory of Lithospheric and Environmental Coevolution, University of Science and Technology of China, Hefei 230026, Anhui, China\\
\small $^{6}$Astronomy \& Astrophysics Section, School of Cosmic Physics, Dublin Institute for Advanced Studies, 31 Fitzwilliam Place, Dublin D02 XF86, Ireland\\
\small $^{7}$Space Science Institute, 4765 Walnut Street, Suite B, Boulder, CO, USA\\[0.5em]
\small Corresponding author: \href{mailto:tom.briand@lam.fr}{tom.briand@lam.fr}
}}

\date{\small Received: 28 March 2026 \quad Accepted: 6 August 2026\\[0.25em]
\href{https://doi.org/10.1051/0004-6361/202660770}{doi:10.1051/0004-6361/202660770}}

\begin{document}
\maketitle

\begin{abstract}
\noindent The formation of Uranus and Neptune remains poorly constrained largely due to uncertain deep elemental abundances. Carbon monoxide (CO), a disequilibrium species in the upper troposphere, provides an indirect constraint on the deep oxygen abundance. We investigate the deep O/H and C/O ratios of the ice giants and their formation history by accounting for meridional variations in atmospheric structure and uncertainties in chemical kinetics. We extended a 1D thermochemical and diffusion model into a pseudo-2D model by including latitudinal variations in key model parameters. The O/H ratio was inferred by matching the modeled upper-tropospheric CO mole fractions to observations, and combined with the deep carbon abundance to derive the C/O ratio. For Uranus, varying tropospheric methane alone yielded O/H $\sim [47-57] \times$ protosolar, whereas allowing $K_{zz}$ to vary expanded the range to O/H $\sim [62-177] \times$ protosolar. For Neptune, the corresponding ranges are O/H $\sim [182-215] \times$ protosolar and O/H $\sim [222-342] \times$ protosolar, respectively, supporting lower oxygen enrichment in Uranus than Neptune. Chemical-network uncertainties have a more modest effect, amounting to at most $\sim 10\%$ of the retrieved oxygen ranges, comparable to the uncertainty associated with the measured CO abundance on Neptune, but larger on Uranus. We computed C/O latitudinal ranges and found C/O $\sim [0.06-0.52]$ on Uranus and $\sim [0.02-0.12]$ on Neptune. Comparison with a protoplanetary disk model suggests different formation or evolutionary pathways for the two planets. Our results highlight the dominant role of vertical mixing in constraining deep oxygen abundance and the importance of accounting for meridional variability and chemical uncertainties. This work also provides a framework for selecting the entry latitude of a future Uranus Orbiter and Probe mission.

\end{abstract}


 \section{Introduction}
Uranus and Neptune remain the least explored planets in the Solar System and were each visited only once by Voyager 2 in 1986 and 1989, respectively. Their long heliocentric distances and low temperatures severely limit remote-sensing capabilities. In particular, the atmospheric opacity from clouds and gases confines infrared spectroscopic observations to the upper atmosphere, down to a few bars, while radio observations can probe down to hundreds of bars \citep[e.g.,][]{Fletcher2020,Hofstadter2024}. As a result, the bulk chemical composition of the ice giants remains essentially unconstrained.
This limitation is critical, as planets of comparable size dominate the exoplanet population \citep[see, e.g.,][]{Atreya2020,deleuil2020}. Understanding the formation and composition of Uranus and Neptune is therefore key to interpreting a large fraction of planetary systems. Their origins remain debated, however. Competing formation pathways predict fundamentally different heavy-element enrichments and compositional signatures.
The two leading scenarios, disk gravitational instability and core accretion \citep[e.g.,][]{Helled2014,Paardekooper2018,Raymond2022}, lead to distinct outcomes. Gravitational instability implies nearly uniform enrichment across elements \citep{Boss2002}, whereas core accretion produces compositions that depend sensitively on how volatiles condensed and were incorporated into the growing core. In this framework, two end-member processes have been proposed: trapping in amorphous ices \citep{Owen1999}, and incorporation into clathrate hydrates \citep{Lunine1985,Mousis2010}. Therefore, it is essential to measure the deep atmospheric abundances of key ice-forming elements because it directly constrains the condensation pathways and physical conditions that shaped the formation of Uranus and Neptune.

An atmospheric entry probe as part of the Uranus Orbiter and Probe (UOP) mission would likely sample the atmosphere down to about 10 bars \citep{Mousis2018}. While this pressure level enables measurements beneath the main cloud deck in the upper troposphere, it remains well above the water cloud, which is expected to reside at depths corresponding to pressures up to several hundred bars, depending on the assumed water abundance and atmospheric density structure \citep{Cavalie2017,Hueso2019,Atreya2020}. Consequently, the deep water abundance must be inferred indirectly, using models that link upper tropospheric measurements of disequilibrium species to thermochemical processes occurring at greater depths.

In the deep troposphere, at pressures greater than thousands of bars, the high temperature allows kinetics to be fast enough to maintain thermochemical equilibrium. The characteristic chemical timescales are therefore lower than that of atmospheric mixing ($\tau_{\mathrm{chem}}<\tau_{\mathrm{mix}}$). As convection transports the gas mass to higher altitudes, it cools down and expands. Eventually, the temperature becomes too low and the kinetics too slow for chemical equilibrium to be maintained. When $\tau_{\mathrm{chem}}=\tau_{\mathrm{mix}}$, the mole fractions of disequilibrium species are frozen, and species are quenched in the upper troposphere, where convective transport dominates \citep{Prinn1977}. CO is a quenched species observed in the upper troposphere of the ice giants \citep{Marten1993,Cavalie2026} and is in deep thermochemical equilibrium with H\textsubscript{2}O \citep{Prinn1977,Prinn1981}. CO has been observed in the stratosphere of both planets, part of which is thought to have an exogenous origin (see \citealt{Teanby2013,Cavalie2014} for Uranus and \citealt{Lellouch2005,Luszcz2013} for Neptune). If most of the CO observed in the troposphere of the ice giants is inherited from the deeper layers, thermochemical models can be used to infer the deep atomic oxygen abundance required to match the CO measurement in the upper troposphere (see \citealt{Cavalie2020} for a review).
The use of thermochemical models requires several input parameters that we list below.

$\bullet$ A temperature profile extrapolated from observable levels to the deep troposphere. Dry or wet adiabats are often used in the giant planets of our Solar System. However, \cite{Guillot1995} initially showed that the condensation of a species such as methane, which is heavier than its environment and highly enriched within the ice giants, can produce a mean molecular mass gradient strong enough to locally inhibit convection. \cite{Leconte2017} proved that if the abundances of the above-mentioned species exceed a certain critical value, this effect can stabilize the atmosphere against double-diffusive convection. This results in energy being transported by radiation at the cloud base, which in turn produces a temperature jump, that is, a sharp thermal vertical gradient whose magnitude depends on the abundances of condensible species. Due to the putative high water and methane abundances on both planets, convection inhibition should be accounted for when modeling the troposphere of the ice giants. \cite{Cavalie2017} showed the impact of using such a three-layer temperature profile (accounting for dry, wet,and radiative processes) in deriving the deep oxygen abundance using a thermochemical and diffusion model in the ice giants.

$\bullet$ Atmospheric chemistry is described in a large network of reactions linking the different species together. Current models adapted to the study of the troposphere of the giant and ice giant planets include C/H/O/N species. The chemical networks of \citet{Venot2012,Venot2015,venot2019reduced,Venot2020, veillet2024} are among the most recent and are validated against experimental data acquired in combustion chambers under conditions relevant to the study of the tropospheres of the giant planets, that is, up to 50 bars and 2400 K. 

$\bullet$ In 1D atmospheric models, vertical mixing is modeled by an eddy diffusion coefficient $K_{zz}$. This coefficient is poorly constrained and is generally estimated to only an order of magnitude using mixing length theory \citep{Stone1976}. However, the rotation of the giant planets can partially suppress convection in their deep atmosphere \citep{Guillot2005}. \cite{Visscher2010} proposed a derivation of $K_{zz}$ as a function of latitude and depth in the giant planets' atmospheres. This was later improved upon and used to constrain the deep water abundance in Jupiter by \cite{Wang2015}.

We extend a 1D thermochemical diffusion model to account for meridional variability in key parameters, including mixing efficiency, temperature structure, and constraints on chemical species from observations and models. We describe the baseline 1D framework and its pseudo 2D extension in Sect. \ref{section:methods}. In Sect. \ref{section:results} we present updated constraints on the deep oxygen abundance in Uranus and Neptune. The uncertainties are quantified by propagating errors in chemical reaction rates.
Combining these results with estimates of the deep carbon abundance, we derived C/O ratios and interpreted them in the framework of protoplanetary disk (PPD) models describing the thermodynamic evolution of trace species \citep{Schneeberger2023,Mousis2024}. These results provide constraints on the formation conditions of Uranus and Neptune, which we present in Sect. \ref{section:formation}. We discuss the various model uncertainties and limitations in Sect. \ref{section:discussion}. Finally, we summarize our main results in Sect. \ref{section:conclusion}.

\section{Thermochemical modeling of the ice giants}
\label{section:methods}
In this section, we build upon the 1D thermochemical diffusion model of \citet{Cavalie2024}, extending it to a pseudo-2D framework. We briefly describe the model and the updates we implemented.

\subsection{Model overview}
The initial condition of this model considers a mixture of H\textsubscript{2}, He, H\textsubscript{2}O, and CH\textsubscript{4}, without chemistry. A simple set of equations links the deep elemental abundances to the mole fractions of the aforementioned species at the upper ($p \sim 2$ bars) and lower ($p \sim 10^5$ bars) boundaries of the model (see eq. (2-4) in \citealt{Cavalie2017}). The measured values of the upper mole fractions of He and CH\textsubscript{4} and of several other parameters we used are summarized in Table \ref{table:constants planets}. Composition, temperature, heat capacities, and mean molecular mass profiles are constructed layer by layer, down to levels where thermochemical equilibrium prevails. Starting this extrapolation just below the measurements at the methane cloud level allows us to ignore complex mean molecular weight effects in the CH\textsubscript{4} cloud (\citealt{Guillot1995,clement2024}), though we account for them in the H\textsubscript{2}O cloud as described next.

\begin{table}[htbp]
\caption{Planet parameters.}
\begin{center}
\begin{tabular}{l r r} 
    \hline \hline
    Planet & Uranus & Neptune \\
    \hline
    $\bar{g_0}$ [m$^2$s$^{-1}$] & 9.01 & 11.27 \\
    $J_2 (\times10^{-6})$ & $3510.68 \pm 0.70$ & $3408.43 \pm 4.50$ \\
    $\Omega [s^{-1}]$ & $1.08\times10^{-4}$ & $1.01\times10^{-4}$  \\
    $F$ [Wm$^{-2}$] & $0.078\pm0.018$ & $0.433\pm0.046$ \\
    $T^{\mathrm{top}}$ [K] & 102.9 & 93.1 \\
    $y_{\mathrm{He}}^{\mathrm{top}}$ & $0.152\pm 0.033$ & ${0.149^{+0.017}_{-0.022}}$ \\
    $y_{\mathrm{CH_4}}^{\mathrm{top}}$ & 0.019-0.044 & 0.018-0.048 \\
    $y_{\mathrm{CO}}^{\mathrm{top}}$ & ${5.6 \pm 0.2 \times10^{-9}}$ & ${0.2\pm0.05 \times 10^{-6}}$ \\
    \hline
\end{tabular}
\end{center}
\tablefoot{Mean surface gravity, computed from $GM/R^2$ (\citealt{jacobson2009,jacobson2014,archinal2018}), second-order gravitational moment (\citealt{jacobson2009,jacobson2014}), rotation rate computed from $2\pi/P_{rot}$ \citep{lecacheux1993,desch1986}, internal heat flux \citep{wang2025internal,pearl1991albedo}, 2 bar temperature measurements \citep{Orton2014,Lindal1992}, and upper tropospheric mole fraction of \ce{He} \citep{Conrath1987,Burgdorf2003}, \ce{CH4}(\citealt{Sromovsky2014,Irwin2021}, latitudinal range) and CO \citealt{Cavalie2026,Luszcz2013,Moreno2017}).}
\label{table:constants planets}
\end{table}

We applied the three-layer profiles of \cite{Leconte2017}, starting from the deeper layers and working upward. In regions in which the water mole fraction reached its saturation pressure curve, a moist adiabat was applied. Otherwise (i.e., below the water cloud), we applied a dry adiabat. When the convection inhibition criterion was satisfied in the water-condensation region, a radiative gradient was applied. 

The thermal profiles are then interpolated on a pressure grid with a more limited sampling ($\sim$ 60 levels) to make the following computations faster. We ensured that the radiative and quenching regions were resolved. The atmospheric composition is calculated at thermochemical equilibrium by Gibbs energy minimization \citep{Agundez2014}, for a given set of species in a chemical network. The reaction rates are computed for every forward reaction, which are then reversed. A constant $K_{zz}$ value is selected. For every species  $i$ at each point of the grid $j$, the model solves the continuity equation, coupling the dynamics and the chemical reactions \citep{venot2012photochimie}, 

\begin{equation}\label{eq:continuity equation}
        \frac{\partial y_i^j}{\partial t} = \frac{1}{n^j} \left(P_i^j - y_i^jn^jL_i^j - div(\overrightarrow{\Phi_i^j}) \right)  ,
\end{equation}
where $y$ is the mole fraction, $t$ is the time, $n$ is the concentration, $P$ and $L$ are the production and loss rates of the species, respectively, and $\overrightarrow{\Phi}$ is the vertical transport flux. We integrated this equation until a steady state was reached. 

We adapted this framework to allow several parameters to vary with latitude. We varied the deep oxygen abundance to find an upper tropospheric CO mole fraction that matches the measurements made on both planets, as described in the following section. We also explored how the uncertainties of the network reaction rates affect the retrieved deep oxygen abundance. In the present work, we assumed our variables are symmetric about the equator. 

\subsection{Upper tropospheric CO}
\label{subsect:CO measurement}
In Neptune, CO was initially detected by \cite{Marten1993}. \cite{Lellouch2005} first demonstrated that the CO profile had an internal component, caused by a deep tropospheric source, and an external component, most likely resulting from an old comet impact. Using interferometric observations in the millimeter wavelength range, \cite{Luszcz2013} and \cite{Moreno2017} constrained the tropospheric CO mole fraction to $0.20\pm0.05$ ppm. This value is also in agreement with Herschel/SPIRE observations \citep{Teanby2019}. For reference, we took 25$^{\circ}$S as a broad estimate of the sub-observer latitude at the time of recent Neptune observations \citep{Irwin2016,Carrion2023}.

\begin{figure}[h]
    \centering
    \includegraphics[width=0.7\linewidth]{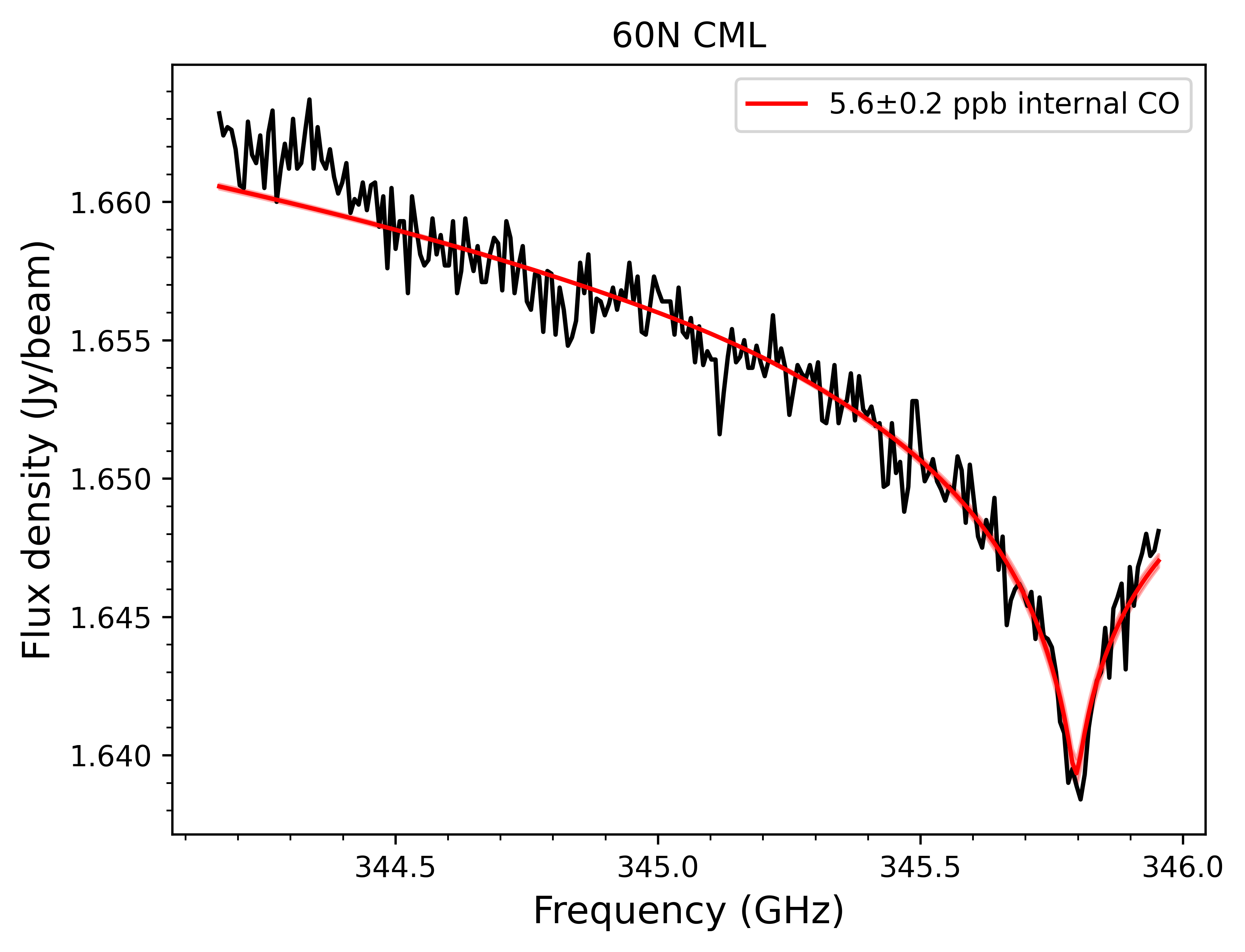}
    \caption{ALMA observations of Uranus' CO (3-2) line. Observations were performed on August 19 and October 18, 2022, as part of project 2021.1.01034.S (PI: S. Luszcz-Cook). Sub-observer latitudes range from 60.2$^{\circ}$ to 61.3$^{\circ}$ between the two observing dates. The best-fit internal CO mole fraction that matches the spectra recorded at the central meridian longitude and 60$^{\circ}$ is $y_{CO}^{top}=5.6 \pm 0.2\times10^{-9}$.}
    \label{fig:CO_measurement_ALMA}
\end{figure}

In Uranus, CO was detected by \cite{Encrenaz2004}. Despite spectrally resolved observations with Herschel, \cite{Teanby2013} and \cite{Cavalie2014} could only demonstrate the existence of an external source of CO. Only recent ALMA observations enabled \citet{Cavalie2026} to unambiguously detect CO in the troposphere of Uranus. They combined observations made in 2022 and 2024 to constrain the upper tropospheric mole fraction to 5.8$^{+0.3}_{-0.3}$\,ppb. 
We used latitude-dependent $y_{\mathrm{CH}_4}^{\mathrm{top}}$ and $K_{zz}$ to constrain the deep O/H ratio. Because the sub-observer latitude varied by almost 10$^{\circ}$ between 2022 and 2024, the $y_{\mathrm{CO}}^{\mathrm{top}}$ derived by \cite{Cavalie2026} covers latitudes between 50$^{\circ}$N and 80$^{\circ}$N. We therefore chose to take only the data from their 2022 dataset (ALMA project 2021.1.01034.S, PI: S. Luszcz-Cook) to avoid mixing data with different observing geometries and to derive an independent $y_{\mathrm{CO}}^{\mathrm{top}}$ value. At the disk-center of the 2022 data, we derived an upper tropospheric CO mole fraction of 5.6$^{+0.2}_{-0.2}$\,ppb (see Fig. \ref{fig:CO_measurement_ALMA}). Owing to the 0.4" beam (with respect to a 3.6" planet), this measurement is made at 60$^{\circ}$ $\pm 10^{\circ}$ (planetocentric latitude) on the central meridian.

\subsection{Temperature extrapolation}
We extrapolated the temperature profile from the upper boundary of the thermochemical--diffusion model, located at 2 bars as mentioned above. This upper tropospheric temperature, denoted as $T^{\mathrm{top}}$, is imposed as a fixed boundary condition and is also independent of the deep oxygen abundance.
To fix $T^{\mathrm{top}}$, we can use the nominal, disk averaged value, given by Spitzer measurements for Uranus ($T^{\mathrm{top}} = 102.9$\,K, \citealt{Orton2014}) and Voyager 2/IRIS occultations for Neptune ($T^{\mathrm{top}} = 93.1$\,K, \citealt{Lindal1992}). 
Recently, \cite{milcareck2024} applied a radiative-convective model to the upper atmosphere of the ice giants. As noted by the authors, the meridional temperature structure found by their model does not match that of IRIS occultations higher up in the atmosphere (i.e., between 0.1 and 0.4 bars). It is unclear whether this discrepancy remains true at deeper levels, as the Voyager 2 temperature structure data is only extrapolated below the tropopause. 

\begin{figure}[h]
    \centering
    \includegraphics[width=0.7\linewidth]{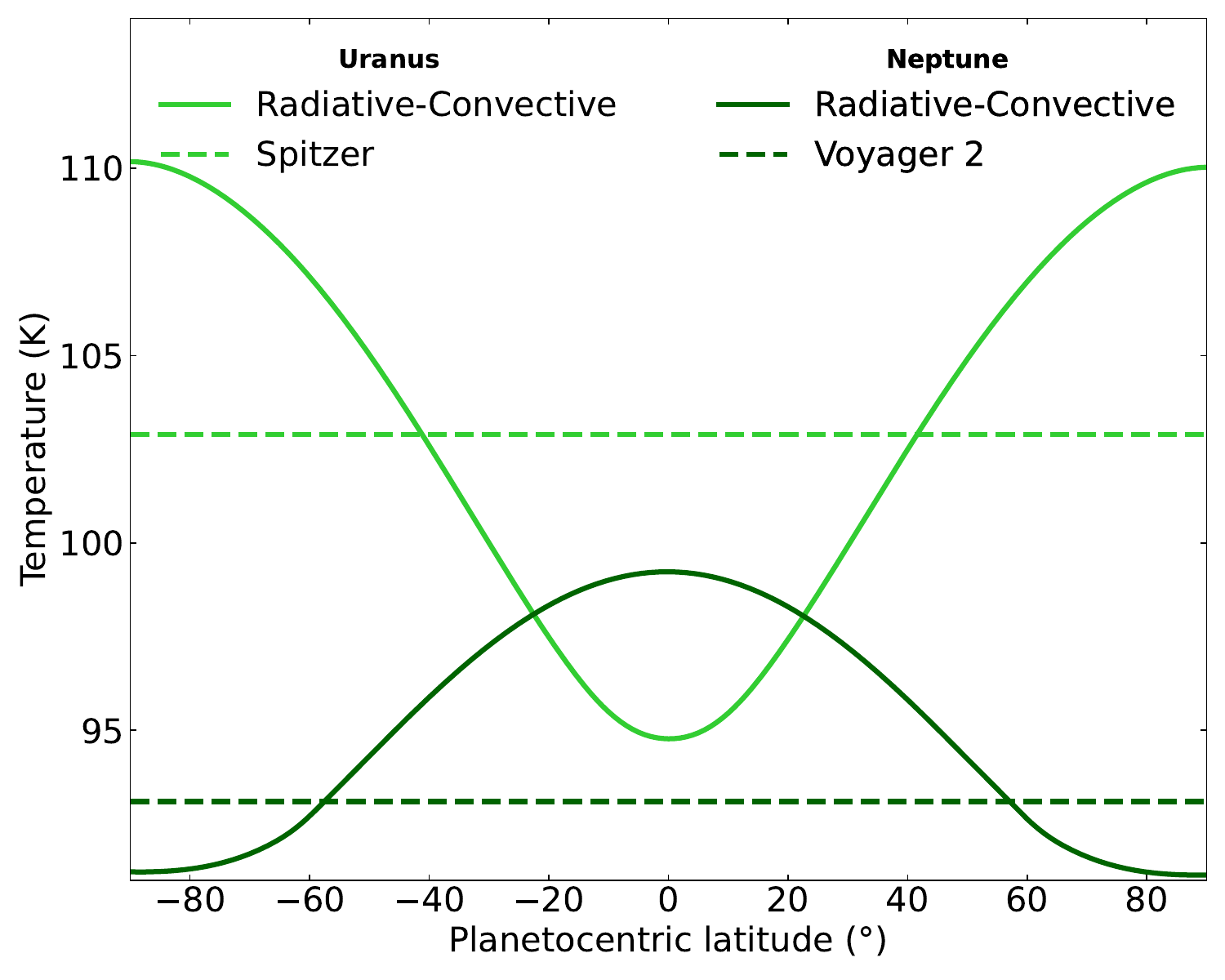}
    \caption{Meridional temperature structure at 2 bars for Uranus (light green) and Neptune (dark green). The dashed lines are disk-averaged measurements (Table \ref{table:constants planets}), and the solid lines result from the radiative-convective model of \cite{milcareck2024}, seasonally averaged. These temperatures were imposed as a boundary condition and were held fixed when O/H was varied.}
    \label{fig:temp2bars}
\end{figure}

We performed simulations using either (i) a constant 2-bar temperature structure across all latitudes and taken from the values recorded by Spitzer and Voyager 2, or (ii) the  meridional temperature structure predicted by the independent radiative-convective model of \cite{milcareck2024}. Both are displayed in Fig. \ref{fig:temp2bars}. The radiative-convective profiles remain consistent with the disk-averaged temperature measurements, but this model predicts a 15\,K increase from the equator to the pole for Uranus and a 10\,K decline from the equator to the pole for Neptune. This model-data discrepancy also seems to be confirmed by recent JWST/MIRI and NIRSpec Uranus observations \citep{roman2025temperature}, which suggests a meridional thermal gradient of 1-2\,K near 2 bars. The simulations we ran adopting these latitudinal upper temperature variations are thus extreme cases, and are used to assess the sensitivity of the inferred O/H abundance to a relatively large meridional variation of the upper tropospheric temperature. In Sect. \ref{section:results} we show that this maximizes the range of oxygen abundances we found on Neptune, and is thus a conservative approach in this case. 

\subsection{Upper tropospheric methane}
\label{subsection:methane}
\begin{figure}[h]
   \centering
   \includegraphics[width = 0.7\linewidth]{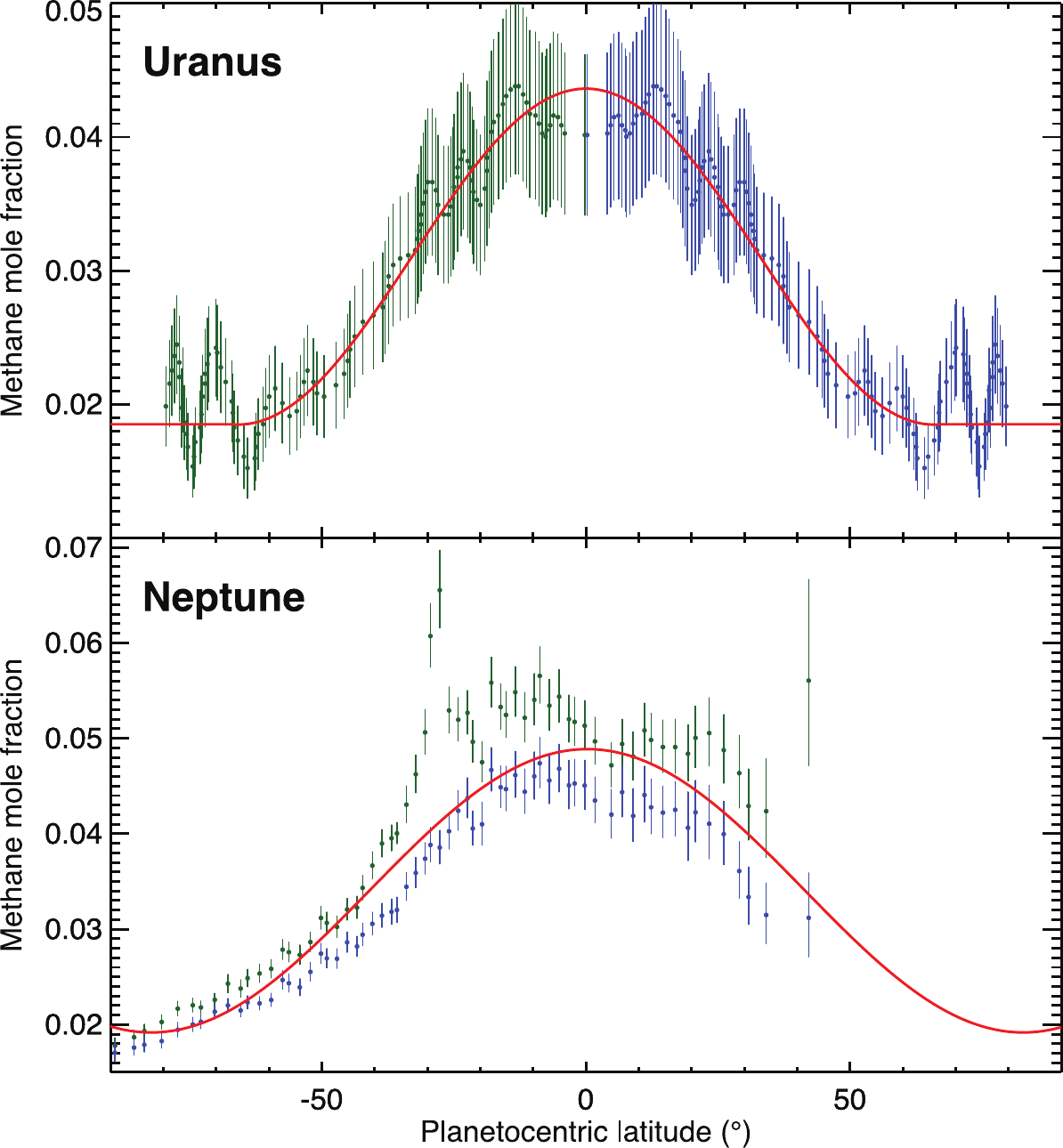}
      \caption{Fits of the upper tropospheric methane mole fraction as a function of latitude. For Uranus (\textit{top}), we fitted the data of \cite{Sromovsky2014}. We show 113 data points between 4°N and 80°N in blue. These values were mirrored in the southern hemisphere. For Neptune (\textit{bottom}), we used the profiles of \cite{Irwin2021}. We present 80 data points from their two inversion models reproc model 1 in green and reproc model 2 in blue.}
      \label{Fig:Fig_methane}
   \end{figure}

The nominal upper tropospheric methane mole fraction used by \cite{Cavalie2017} and \cite{Venot2020} is $y^{\mathrm{top}}_{\mathrm{CH}_4} = 0.04$ on both planets. Spatially resolved observations on both planets indicate that methane is depleted toward the poles in the upper troposphere. The profiles we used are shown in Fig. \ref{Fig:Fig_methane}. For Uranus, we used the meridional profile of \cite{Sromovsky2014}, based on HST/STIS observations made in 2012 between 300 and 1020 nm. We assumed a 15\% uncertainty for each value. For latitudes north of 67$^{\circ}$N, we fixed the methane mole fraction at its minimum value ($y^{\mathrm{top}}_{\mathrm{CH}_4} = 0.019$). For Neptune, we used data from \cite{Irwin2021}, based on Multi Unit Spectroscopic Explorer (MUSE) visible / near-infrared observations performed in 2019. We fitted data from their reproc model 1 and reproc model 2 simultaneously. These two inversion models differ in the assumed cloud scattering properties as a function of latitude and yield differences of about 0.01--0.02 in methane mole fraction. The maximum fitted mole fraction of methane on both planets is reached at the equator, being slightly higher on Neptune ($y^{\mathrm{top}}_{\mathrm{CH}_4} = 0.048$) than on Uranus ($y^{\mathrm{top}}_{\mathrm{CH}_4} = 0.042$). 

\subsection{Eddy diffusion coefficient} \label{subsect:kzz}
Previous works (\citealt{Cavalie2017}, \citealt{Venot2020}) used a constant value of the eddy diffusion coefficient, set at $K_{zz} = 10^8$ cm\textsuperscript{2}s\textsuperscript{-1}, as predicted by the mixing length theory (MLT) within an order of magnitude uncertainty \citep{Stone1976}.
Within the MLT framework, $K_{zz}$ is proportional to the product of the vertical convective velocity $w$ and the characteristic mixing length scale $L$. It can therefore be expressed as
\begin{equation}
    K_{zz} \sim wL\sim\left(\frac{\alpha g F}{\rho c_p}L\right)^{1/3}L
    \label{eq:Kzz MLT},
\end{equation}
where $\alpha$ is the thermal expansion coefficient [K$^{-1}$], $g$ is the acceleration of gravity [m s$^{-2}$], $F$ is the internal heat flux [W m$^{-2}$], $\rho$ is the density [kg m$^{-3}$] and $c_p$ is the heat capacity per unit mass at constant pressure [J kg$^{-1}$ K$^{-1}$]. This formulation does not account for the fast rotation of the giant planets which can partially suppress convection \citep{Stevenson1979}. A new latitude-dependent formulation was proposed by \cite{Visscher2010} but it is only valid for latitudes far from the equator. \cite{Wang2015} proposed a formulation valid for non-equatorial latitudes, which we summarize hereafter. The Coriolis force strongly affects convective motion under rotation, and in the limit of low Rossby number, constrains them to columnar-like flow due to the Taylor-Proudman theorem \citep{Guillot2005}. Thus, \cite{Wang2015} proposed that the convective overturn timescale should be limited by the rotation rate of the planet such that $\tau\sim1/\Omega$, resulting in

\begin{equation}
    K_{zz} \sim wL\sim\frac{\alpha g F}{\rho c_p\Omega^2}
    \label{eq:Kzz scale}
.\end{equation}

The scalings are found using laboratory experiments (e.g., \citealt{Fernando1991}) involving a water tank in rotation, heated from the bottom and open at the top to simulate convection. Observables such as temperature or directional velocities are retrieved. These experiments reveal two convective regimes, depending on the rotation rate.
The transition between these two regimes is defined by a value of the natural Rossby number characteristic of the system's rotation rate, $Ro^*_t = 0.015$. \cite{Wang2015} found that this corresponds to a transition latitude $\phi_t$, such that
\begin{equation}
    \sin{\phi_t}=\left(  \frac{\alpha g F}{\rho c_p H^2 {Ro_t^*}^2} \right)^{1/3} \frac{1}{\Omega}
,\end{equation}
where the characteristic length scale chosen is $L=H$, the atmospheric scale height [m]. In practice, \cite{Smith1998} showed that in the troposphere of giant planets, $L$ is better estimated by a fraction of $H$.
The slow-rotation scaling is the same as predicted by the MLT and applies below the transition latitude. Otherwise, a fast-rotation scaling applies. The two regimes of the eddy diffusion coefficient are defined as
\begin{equation}
    \label{eq:Kzz_lowlat}
    K_{zz} = (0.18 \pm 0.02) \left( \frac{\alpha g F}{\rho c_p} \right)^{1/3} H^{4/3} \quad \text{for } \phi < \phi_t
\end{equation}
\begin{equation}
    \label{eq:Kzz_highlat}
    K_{zz} = (50 \pm 10) \frac{\alpha g F}{\rho c_p (\Omega \sin \phi)^2} \quad \text{for } \phi > \phi_t.
\end{equation}

For the latitudinal dependence of $g$, we used the formula from \cite{Guerlet2014}, 

\begin{equation}
    g ({\phi}) = \bar{g_0} \left[ 1 - \frac{3}{4}f + \frac{3}{4}J_2 - \left( 2f-\frac{15}{4}J_2 \right) \cos 
    2\phi\right]
    \label{eq:g by S.Guerlet}
,\end{equation}
where $f=\frac{R_{eq}-R_{pole}}{R_{pole}}$ is the oblateness of the planet with $R_{pole}$, $R_{eq}$  the polar and equatorial radii, respectively, $\bar{g_0} = \frac{GM}{\bar{R}^2}$ is the mean surface gravity with $\bar{R}$ the mean radius, and $J_2$ is the second order gravitational moment. The values of the constant parameters for each planet are given in Table \ref{table:constants planets}. \\

\begin{figure}[h]
    \centering
    \includegraphics[width = 0.49\linewidth]{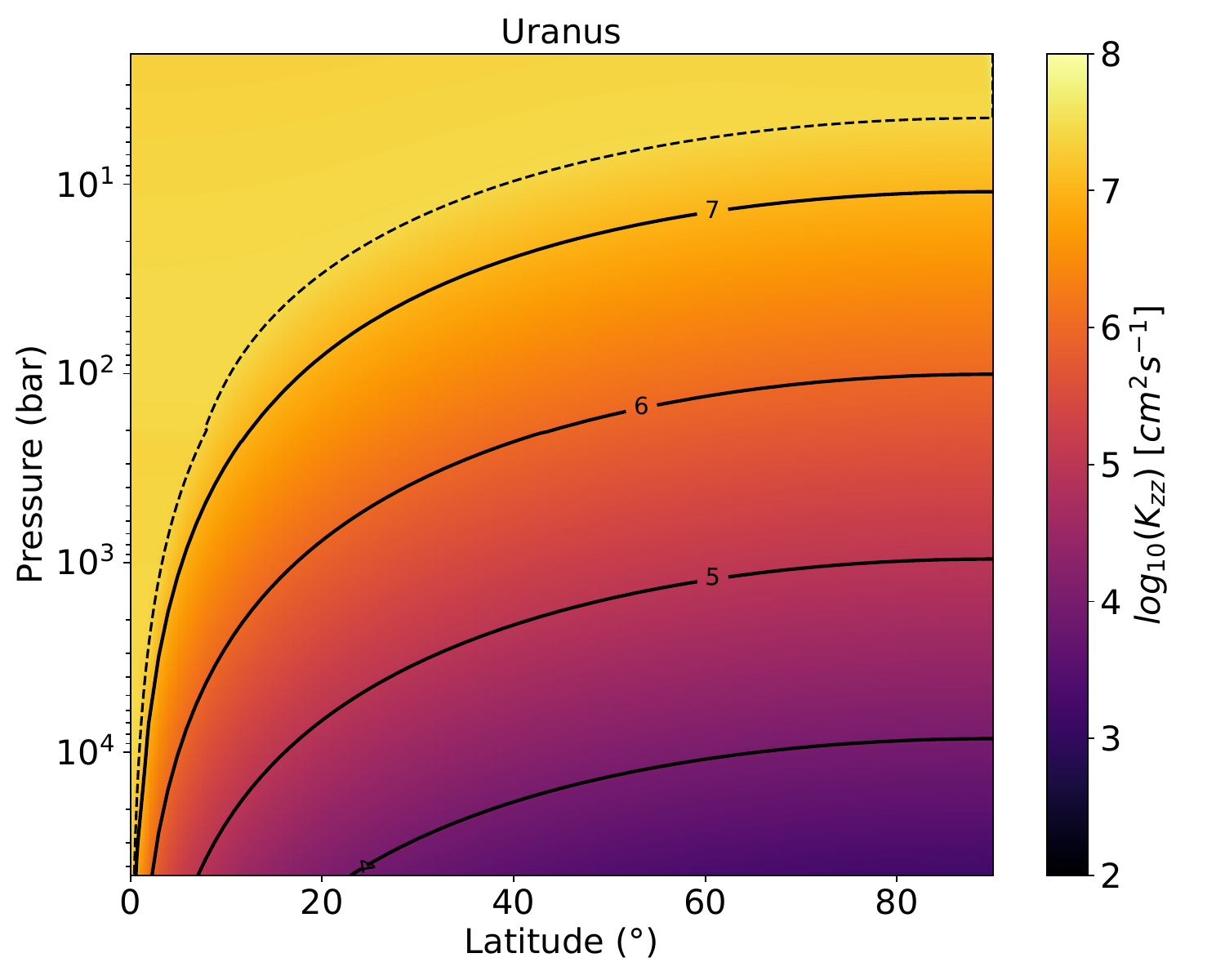} 
    \hfill
    \includegraphics[width = 0.49\linewidth]{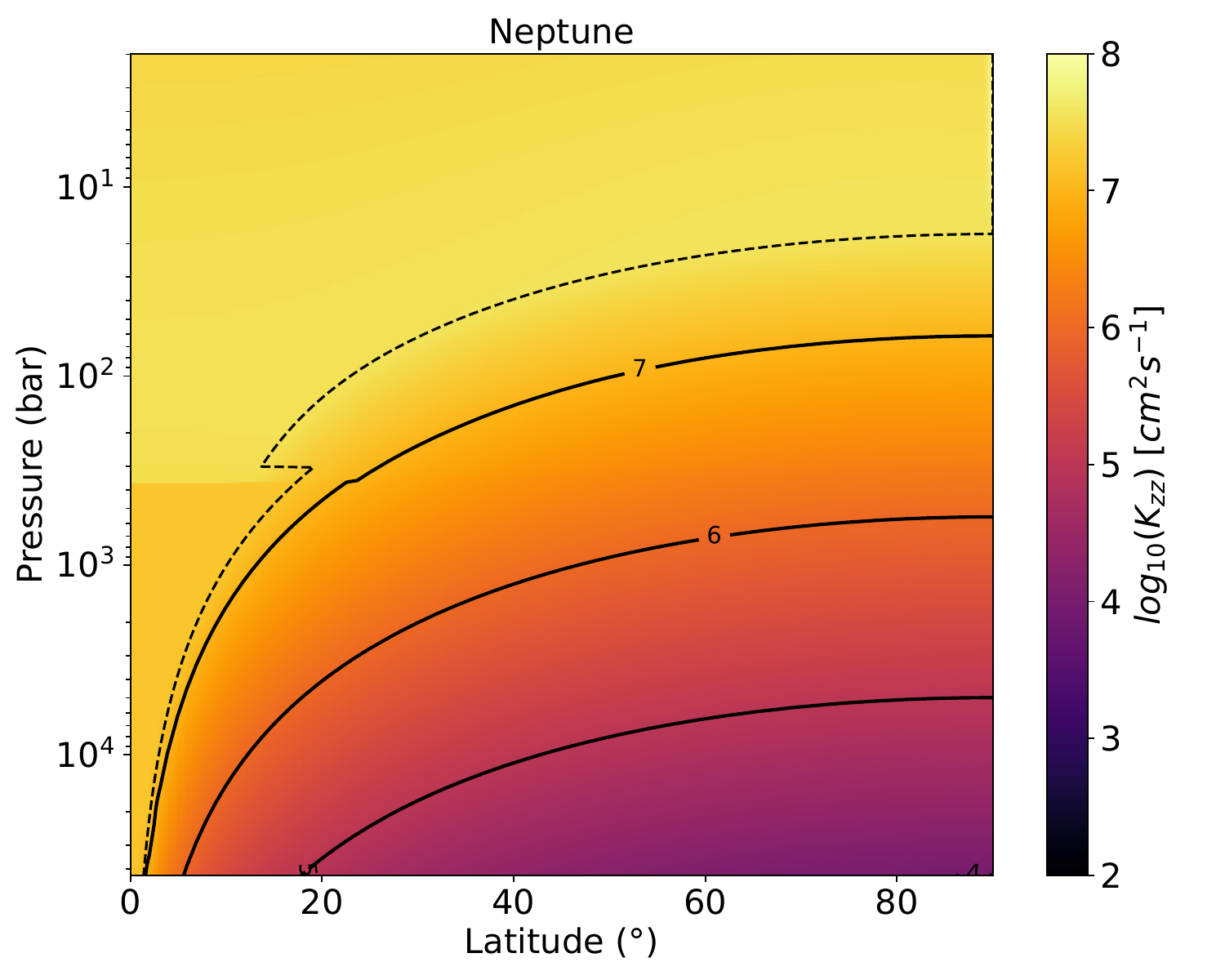}
    \caption{Eddy diffusion coefficient computed on Uranus (\textit{left}) and Neptune (\textit{right}) as a function of pressure and latitude following the prescription of \cite{Wang2015}. The dashed line corresponds to the transition latitude between the fast- and slow-rotating regimes. For each degree of latitude, a different thermal profile was computed with the corresponding value of $y_{CH4}^{top}$ using the fits presented in Sect. \ref{subsection:methane}. This would result in a different deep oxygen abundance required to fit the CO upper tropospheric value depending on latitude. For simplicity, the $K_{zz}$ profiles shown here were computed with the thermal profiles obtained using the deep oxygen abundance derived by \cite{Venot2020} for each planet.}
    \label{fig:Kzz_both}
\end{figure}

The quantities $c_p$ and $H$ vary with pressure and are computed from the temperature profile, which itself varies with latitude. Figure \ref{fig:Kzz_both} shows the resulting eddy diffusion coefficient as a function of pressure and latitude in the troposphere of both planets. $K_{zz}$ generally decreases with depth and latitude. In particular, around the quench level, the atmosphere is better mixed at low latitudes. The value of $K_{zz}$ at low latitudes is the same scaling as predicted by MLT, but due to the prefactor in \eqref{eq:Kzz_lowlat}, the value of $K_{zz}$ at the quench level is always lower than the nominal MLT value on both planets ($K_{zz} = 10^8$ cm$^2$s$^{-1}$).
Below the transition latitude, $K_{zz}$ increases with pressure until reaching the water condensation level, at which point the trend is reversed. We discuss the behavior of $K_{zz}$ in the water condensation region in Sect. \ref{subsect:uncertainty kzz}.

\subsection{Chemical network}

In a chemical network, each reaction links one or more reactants to products through a reaction rate $k$, which can be temperature- and sometimes pressure-dependent. Our knowledge of these rates derives from either experimental measurements or theoretical calculations. The accuracy of thermochemical and photochemical models is therefore limited by the uncertainties associated with these rate constants \citep{Dobrijevic1998, dobrijevic2010neptune, Benne2022}. Knowledge of these uncertainties enables the development of reduced chemical networks, which allow for faster computations, with predictive capability within the overall network uncertainty \citep{Hue2015, Hue2018}. It also allows us to perform a global sensitivity analysis (SA) in order to identify key reactions in chemical networks whose accuracy drastically impacts the predictivity of the chemical network \citep{Dobrijevic2011}.

When the forward rate $k_f$ has been computed, either the reverse rate $k_r$ or the equilibrium constant $K_{eq}$ can be computed from each other using $K_{eq}$ = $k_f$/$k_r$. \cite{Cavalie2017} used the C/H/O/N network of \cite{Venot2012}, which we refer to as V12. We used the full (V20) and reduced (V20r) networks of \cite{Venot2020}, which have been applied to the troposphere of Uranus and Neptune. The V20 network was updated from V12 with refined methanol chemistry, the former results in an O/H solar enrichment that is lower by two to three times than necessary compared to V12 in the ice giants \citep{Venot2020}. V20 contains 108 species and 1906 total reactions, while the reduced network contains 44 species and 582 reactions. We also used the updated C/H/O/N network of \cite{veillet2024} that we denote V24. The latter contains 175 species and 2550 reactions. These are all part of an effort to experimentally validate chemical networks with combustion chambers, and can be found on the EXACT ANR project page \footnote{https://www.anr-exact.cnrs.fr/chemical-schemes/}.

\subsection{Uncertainty propagation}
\label{subsect:uncertainty_propagation}
Solving the continuity equation with a large and nonlinearly coupled chemical network results in significant uncertainty propagation (UP), affecting the resulting mole fractions of most species \citep{Dobrijevic1998}. \cite{venot2019reduced} assessed the impact of the global uncertainty propagation of the reduced V12 model in the troposphere of the warm Neptune GJ 436b. Recently, \cite{agundez2026} carried out a SA on a chemical model applied to several exoplanets. 

Following a similar Monte-Carlo (MC) approach, we quantified the error in the modeled upper-tropospheric CO abundance arising from uncertainties in the reaction rates. The UP is performed with the V20r network. The V20 and V24 networks are used only for comparisons with their nominal reaction rates, that is, their rate uncertainties are not propagated because of their substantially greater computational cost.

For one MC realization $i$, a random coefficient $\epsilon_{ij}$ is sampled from a standard normal distribution (mean 0, standard deviation 1) for every forward reaction $j$ independently. We therefore computed a new reaction rate $k_j'$, from the nominal rate coefficient $k_j$ and the uncertainty factor $F_j$ of a given forward reaction $j$,

\begin{equation}
    k_j'(T) = \exp( \ln k_j(T) + \ln F_j(T)\times \epsilon_{ij} )
.\end{equation}
This is equivalent to randomly drawing $k_j'(T)$ from a lognormal distribution whose median is the nominal reaction rate $k_j$ and whose dispersion is controlled by the magnitude of the uncertainty factor $F_j$.
Within one MC realization, the rate coefficients of all forward reactions are perturbed simultaneously. The modified reverse rates were computed from the perturbed forward rate and the reaction equilibrium constant $K_{i,eq}$, in order to maintain thermochemical equilibrium below the quenching level, following \cite{venot2019reduced}.
For a given latitude and elemental composition, the thermochemical--diffusion model is integrated to steady state for $N_{\mathrm{MC}}$ independently perturbed chemical networks. We subsequently obtained a distribution of the upper tropospheric CO mole fractions, $y_{\mathrm{CO},j}^{\mathrm{top}}$. We can therefore derive a model uncertainty, denoted as $y_{\mathrm{CO}}^{\mathrm{top\pm}}$, calculated as the upper and lower 1$\sigma$ limits of the $y_{\mathrm{CO},j}^{\mathrm{top}}$ distribution. These limits correspond to a set of two reaction rates $k_j'^{\pm}$ produced by two MC iterations. The application of this procedure to the nominal oxygen abundance retrievals, and the conversion of the resulting CO model uncertainty into an uncertainty on O/H are described in the following section.

\section{Results}
\label{section:results}
In this section, we highlight the deep oxygen abundance that fits the upper tropospheric CO mole fraction $y_{CO}^{top}$ measured for both planets (see Table \ref{table:constants planets} and Sect. \ref{subsect:CO measurement}) using our thermochemical and diffusion model. The measured CO values are taken as disk-averaged quantities. Using the V20r network, we retrieved a deep O/H ratio for each degree of latitude using the temperature, CH$_4$ abundance, and eddy diffusion coefficient appropriate for that latitude and the disk-averaged CO abundance. In this way, we found a range of deep O/H ratios which we took to represent the most likely range of the true O/H ratio. Here we express our results with respect to the protosolar oxygen abundance, that is, 8.85 dex \citep{lodders2025solar}.

Every 10$^{\circ}$ of latitude, we then performed N$_{\mathrm{MC}}$ = 1000 UP simulations in which all reaction rates were perturbed simultaneously as described in Sect. \ref{subsect:uncertainty_propagation}. The 1000 realizations were calculated with O/H fixed at the nominal best-fitting value following the procedure detailed above. We thus derived an error bar on the modeled $y_{CO}^{top}$ as detailed in the previous section. We then converted this into an uncertainty on the retrieved deep oxygen abundance. Using the nominal V20r rates, we determined the O/H ratios required by the model to reproduce $y_{\mathrm{CO}}^{\mathrm{top,+}}$ and $y_{\mathrm{CO}}^{\mathrm{top,-}}$.
Another approach would be to (i) retrieve the set of perturbed rate coefficients $k'^{\pm}$ from the pair of simulations that match the $y_{\mathrm{CO}}^{\mathrm{top,\pm}}$, (ii) run the thermochemical and diffusion model using the same set of perturbed rate coefficients $k'^{\pm}$, and (iii) find the corresponding pair of O/H ratios that fits the observed $y_{\mathrm{CO}}^{\mathrm{top}}$. We tested the two methods and found a similar O/H interval. The former is more straightforward and time-efficient and was therefore chosen.

We also compared our results with the enrichment in protosolar oxygen derived from a 1D thermochemical and diffusion model reported by \cite{Venot2020}. We have updated these values using the most recent Solar System abundances of \citet{lodders2025solar}. For Uranus, they found $\mathrm{O/H} < 35 \times \mathrm{protosolar}$ by fitting the upper limit of $y_{\mathrm{CO}}^{\mathrm{top}}< 2.1$ ppb \citep{Teanby2013}. For Neptune, they found $\mathrm{O/H} \sim 190 \times \mathrm{protosolar}$, by fitting the same value $y_{\mathrm{CO}}^{top}$ as this study ($y_{\mathrm{CO}}^{\mathrm{top}} = 0.20$ ppm). \\

\subsection{Uranus}

\begin{figure}[h]
    \centering
    \includegraphics[width = 0.7\linewidth]{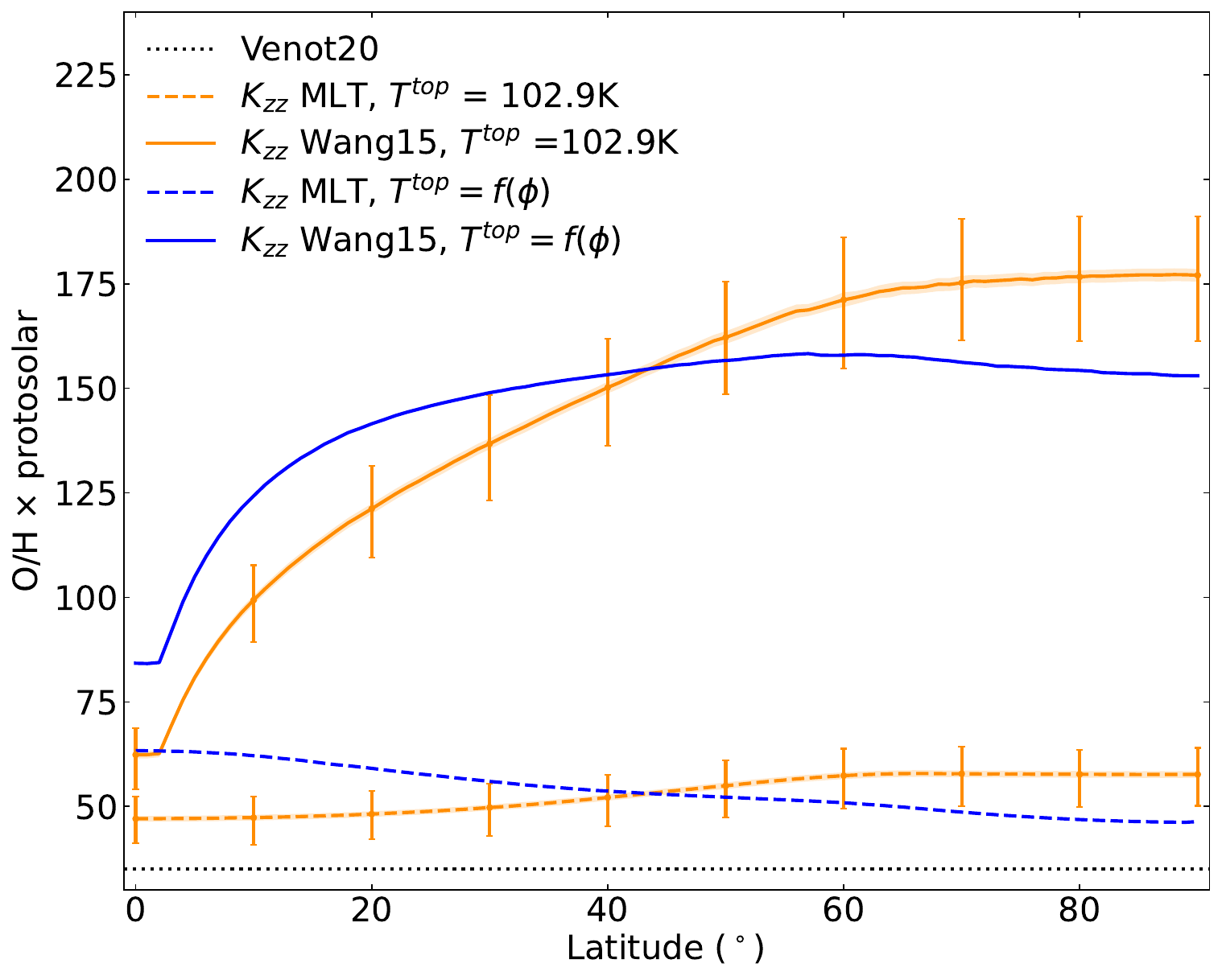}
    \caption{Deep oxygen abundance with respect to the protosolar value as a function of latitude, fitting the upper tropospheric CO mole fraction measurement on Uranus using the V20r chemical scheme. The updated result from \cite{Venot2020} is represented by the dotted dark line (1D model). The $K_{zz}$ profile either varied with latitude and pressure (solid curves) or was fixed at the MLT value (dashed curves). The orange curves correspond to the case where the upper tropospheric temperature was set at the disk-averaged measured value. The error bars were derived every 10° of latitude and correspond to the chemical scheme uncertainties on the CO abundance (see text). The filled orange region represents the deep oxygen abundance obtained when fitting the upper and lower limit of the measured CO mole fraction. The results obtained using a latitudinally varying upper tropospheric temperature are shown with the blue curves for comparison.}
    \label{fig:O_enrich_Ura}
\end{figure}

The best fits of the enrichment with respect to protosolar oxygen with respect to latitude on Uranus are displayed in Fig. \ref{fig:O_enrich_Ura}. When $K_{zz}$ is set constant with latitude at the MLT value ($K_{zz} = 10^8$ cm$^2$\,s$^{-1}$), and only the upper tropospheric methane is allowed to vary, while $T^\mathrm{top} = 102.9$ K, we found $\mathrm{O/H} \sim [47-57] \times \mathrm{protosolar}$. We found a similar range when $T^{\mathrm{top}}$ was allowed to vary with latitude as predicted by a radiative-convective model. In the latter, the temperature is minimum at the equator and maximum at the pole and $y_{\mathrm{CH_4}}^{top}$ follows the opposite trend. As explained in \cite{Cavalie2017}, a higher $y^{\mathrm{top}}_{\mathrm{CH}_4}$ value results in a deeper condensation level and a lower deep tropospheric temperature. These two effects create diverging temperature extrapolation profiles that result in a $\sim$100 K difference between the equator and the poles around the quench level for similar enrichment values. Thereby, when $T^{\mathrm{top}}$ is allowed to vary, the quench level is located higher in the troposphere at low latitudes, requiring a higher O/H enrichment to fit the observed CO mole fraction.

\begin{figure}[t]
    \centering
    \includegraphics[width = 0.49\linewidth]{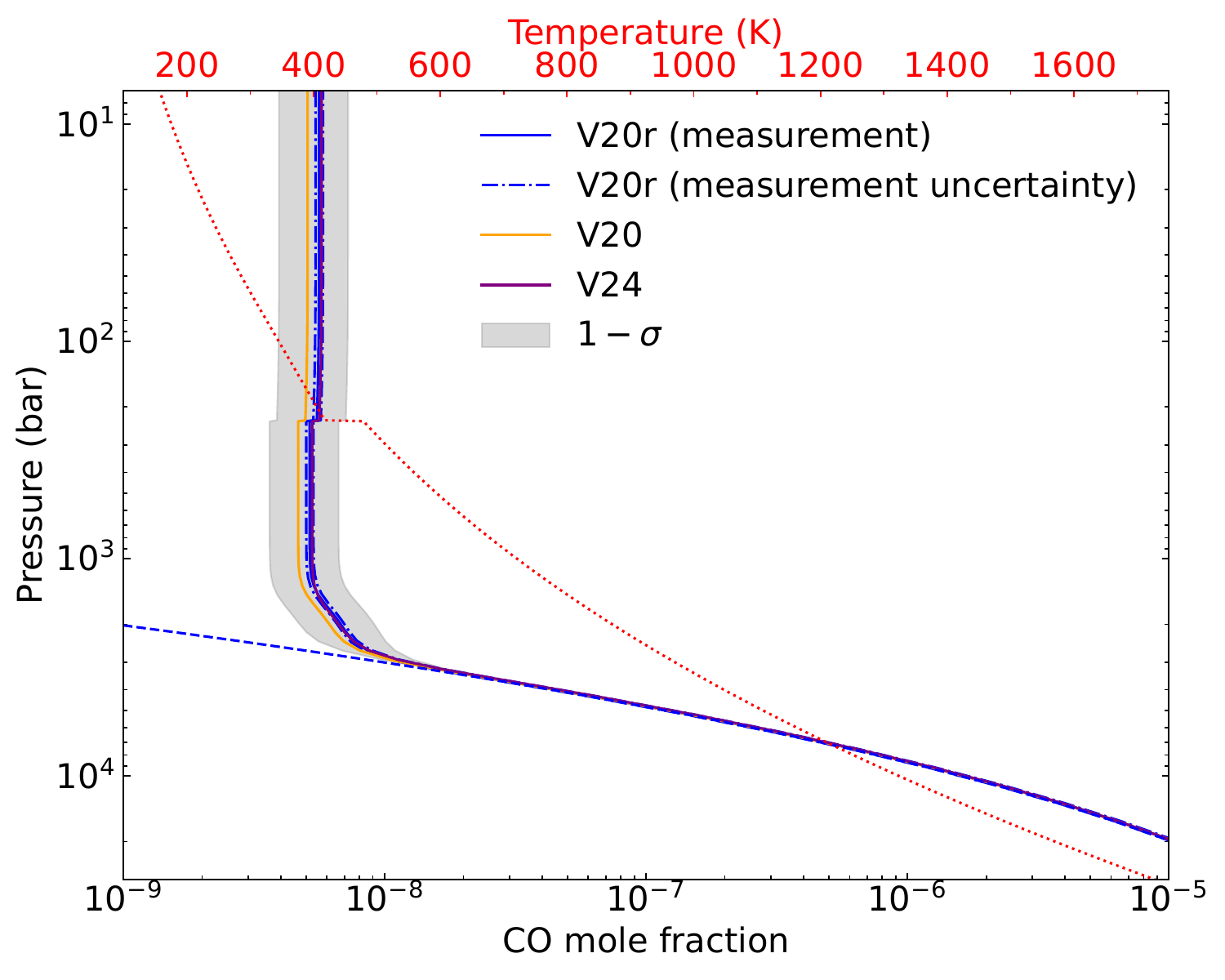}
    \hfill
    \includegraphics[width = 0.49\linewidth]{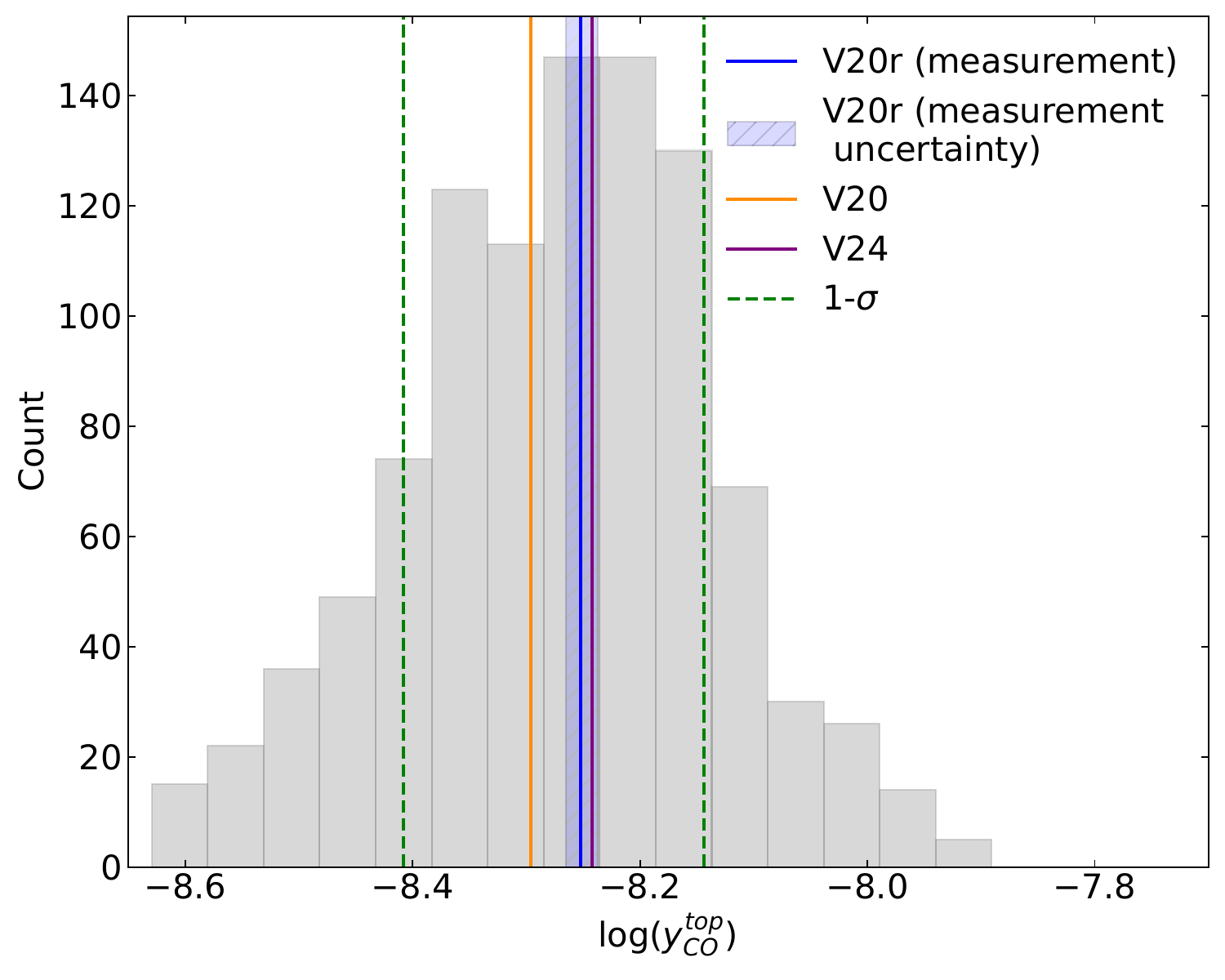}
    \caption{\textit{Top:} Equatorial CO vertical profiles targeting the upper tropospheric measurement on Uranus for the V20r network (solid blue) and with the  $K_{zz}$ formulation of \cite{Wang2015}. The results using the V20 (orange) and V24 (purple) networks at the same oxygen enrichment are also shown for comparison. The dashed lines are set at chemical equilibrium. 
    The dash-dotted blue curves are the profiles targeting the $\pm1$$\sigma$ interval around the $y_{CO}^{top}$ measurement with the V20r network. The gray region represents the 1$\sigma$ V20r network uncertainties, obtained after 1000 MC simulations.
    The matching temperature profile is also shown in red.
    \textit{Bottom:} The corresponding distribution of the retrieved CO mole fraction at the top of our atmosphere after the 1000 MC runs. The solid lines are the upper CO mole fraction value of the profiles shown in the top panel. The transparent blue region correspond to the range of the measurement uncertainty. The dashed green lines correspond to 1$\sigma$ of the distribution.}
    \label{fig:CO_profiles_Ura}
\end{figure}

When varying $K_{zz}$ with latitude, the deep oxygen abundance ranges from 62 to $177\times \mathrm{protosolar}$. The transition region between the fast-rotating and slow-rotating $K_{zz}$ is located at $ \phi_t \sim 5^{\circ}$ at the quench level. There is a factor of 3 difference between the O/H computed at the pole using the MLT and the latitude dependent formulation of $K_{zz}$. 
This highlights that the eddy diffusion coefficient is, as expected, the key parameter in our model and that convection inhibition due to planetary rotation is important for modeling the deep troposphere and is further discussed in Sect. \ref{subsect:uncertainty kzz}. 

Interestingly, going from the mid-to-high latitudes, the competing effects of increasing $T^{\mathrm{top}}$ and reducing $K_{zz}$ lead to a plateau in the deep oxygen enrichment from the mid-latitudes to the poles. In this case, we found  $ \mathrm{O/H} \sim [84-153] \times  \mathrm{protosolar}$. This will be addressed in Sect. \ref{section:discussion}.

Because radiative-convective modeling of the $T^{\mathrm{top}}$ variation remains challenging \citep{milcareck2024}, we only display the results of the uncertainty propagation using the V20r network and a constant $T^{\mathrm{top}}$. The result on the upper tropospheric CO abundance is shown as an example at the equator in Fig. \ref{fig:CO_profiles_Ura}. We obtained a lower uncertainty value of $y_{\mathrm{CO}}^{\mathrm{top},-}=3.9\times10^{-9}$ and an upper uncertainty value of $y_{\mathrm{CO}}^{\mathrm{top},+}=7.2\times10^{-9}$. The uncertainties induced by this larger chemical network are thus greater than the uncertainty on the measured CO abundance on Uranus ($y_{\mathrm{CO}}^{\mathrm{top}}=5.6 \pm 0.2\times10^{-9}$.
V20 and V24 both agree well with V20r ($\sim$ 10\% and $\sim$ 5\% relative error on $y_{\mathrm{CO}}^{\mathrm{top}}$, respectively) and are well within the V20r network 1$\sigma$ uncertainty. This justifies the use of the reduced network in our study, as V20r saves significant computational time (1000 UP runs require $\sim$40 hours instead of $>$6 days at high latitudes). We found that the uncertainty of the chemical scheme results in a $\sim 10\%$ alteration of the deep oxygen protosolar enrichment.  
Finally, we report here the O/H enrichment near the sub-observer Uranus latitude at which the 2022 and 2024 ALMA data were recorded \citep{Cavalie2026}, that is, ($\phi\sim 60 ^{\circ}$). Using a fixed $T^{\mathrm{top}}$, we obtained $ \mathrm{O/H} \sim 56_{-7}^{+7} \times$ protosolar with a constant $K_{zz}$, and $\mathrm{O/H} \sim 171_{-16}^{+15} \times$ protosolar otherwise. This difference will be discussed in Sect. \ref{subsect:pseudo2D limits}.

\subsection{Neptune}

\begin{figure}[h]
    \centering
    \includegraphics[width = 0.7\linewidth] 
    {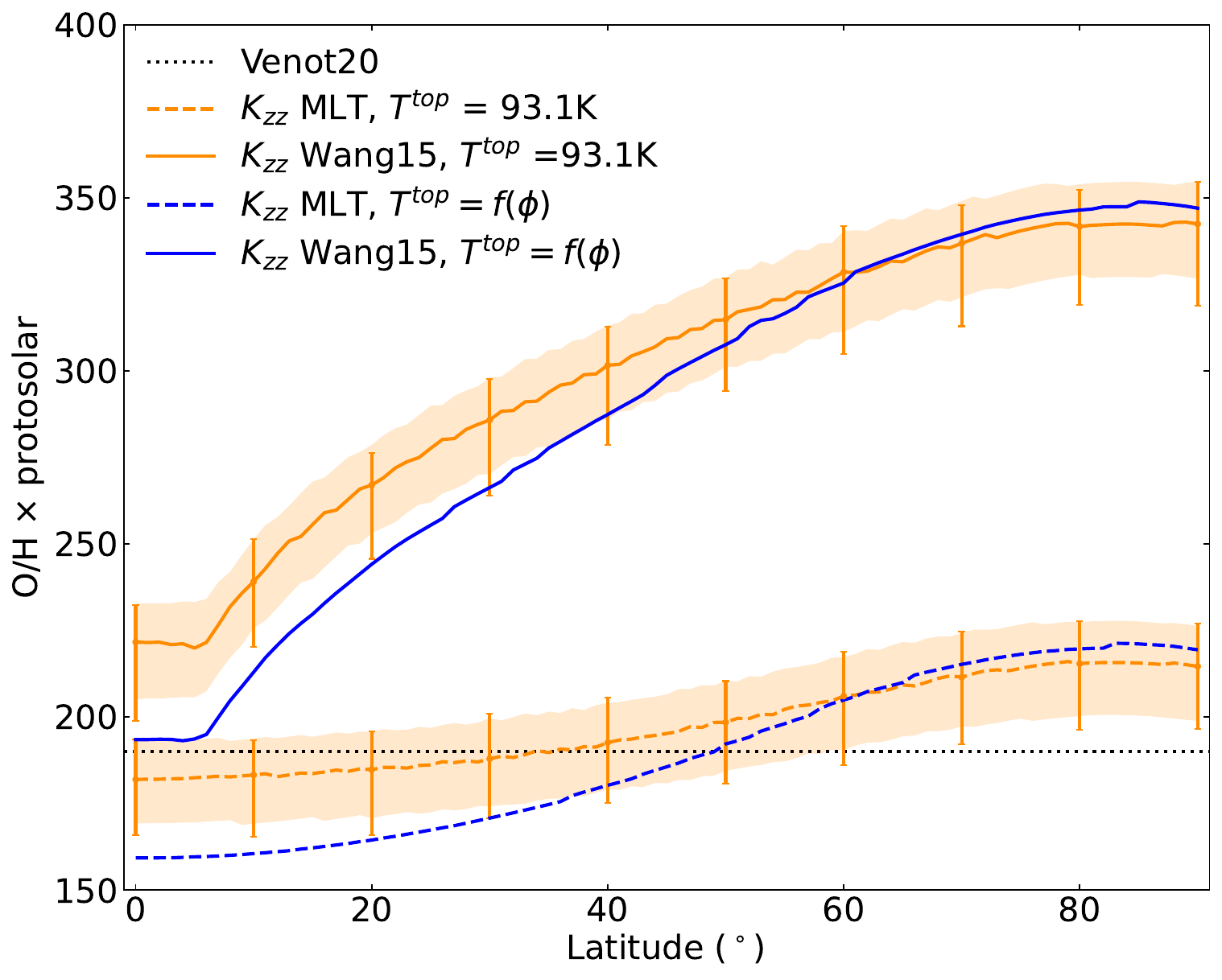}
    \caption{Same as Fig. \ref{fig:O_enrich_Ura}, but for Neptune.}
    \label{fig:O_enrich_Nep}
\end{figure}

The best fits for the enrichment with respect to protosolar oxygen with respect to latitude on Neptune are displayed in Fig. \ref{fig:O_enrich_Nep}. When $K_{zz}$ and $T^{\mathrm{top}}$ are both constant with latitude, we found  O/H$ \sim [182-215] \times \mathrm{protosolar}$ with the maximum being reached at near polar latitudes. When $T^{\mathrm{top}}$ varies, the same trend is found with increased meridional variation as $\mathrm{O/H} \sim [159-219]\times \mathrm{protosolar}$. In the Uranus case, at the MLT value, we noted an inverse trend when $T^{\mathrm{top}}$ varies. On Neptune, the radiative-convective model indeed shows that $T^{\mathrm{top}}$ decreases toward the poles, and the methane mole fraction follows the same trend. As previously mentioned, a lower methane abundance favors an increase in the deep tropospheric temperature, but a lower $T^{\mathrm{top}}$ clearly results in a lower deep tropospheric temperature. Therefore, the two effects compete, and the resulting latitudinal difference in the temperature is lower than on Uranus.  

When we varied $K_{zz}$ with latitude, we derived $\mathrm{O/H} \sim [222-342] \times \mathrm{protosolar}$ and $\mathrm{O/H} \sim [193-349] \times \mathrm{protosolar}$ when $T^{\mathrm{top}}$ was not fixed. For Neptune, the transition region between the fast-rotating and slow-rotating $K_{zz}$ regimes is located farther from the equator than for Uranus, at $ \phi_t \sim 10^{\circ}$ around the quench level.

\begin{figure}[h]
    \centering
    \includegraphics[width = 0.49\linewidth]{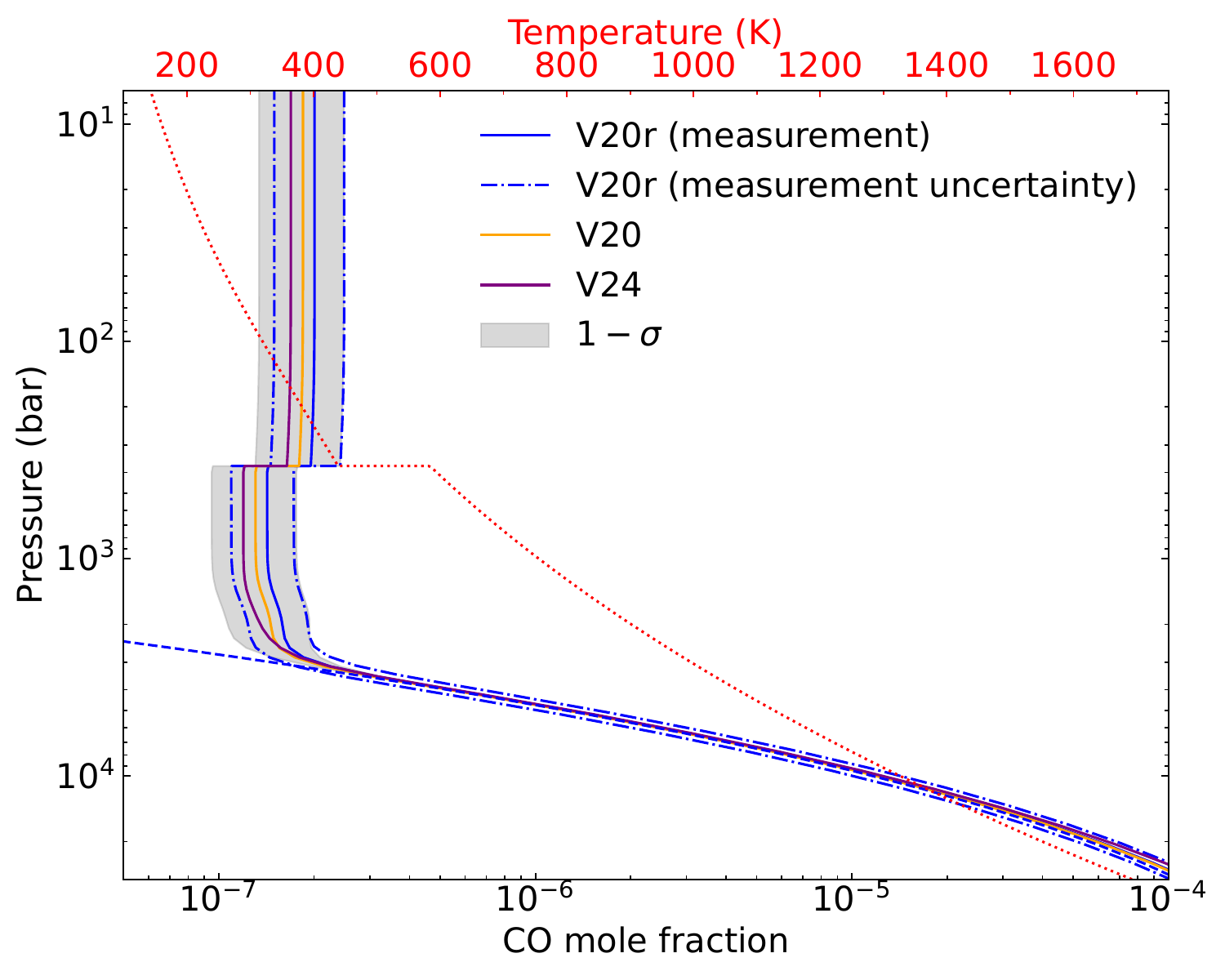}
    \includegraphics[width = 0.49\linewidth]{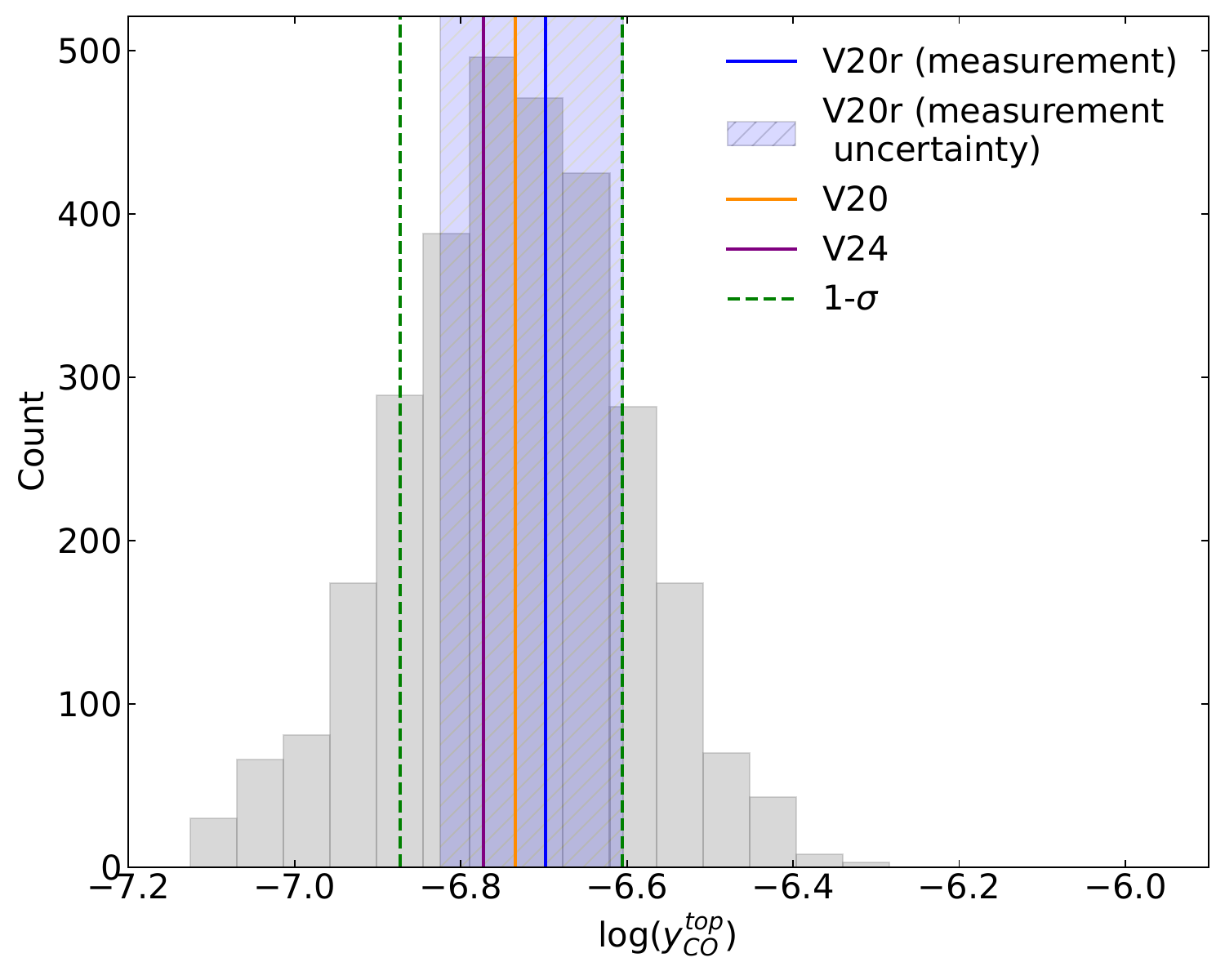}
    \caption{Same as Fig. \ref{fig:CO_profiles_Ura}, but for Neptune.}
    \label{fig:CO_profiles_Nep}
\end{figure}

The resulting UP with the CO profile at the equator is shown in Fig. \ref{fig:CO_profiles_Nep}. At the equator, we found a lower value of $y_{\mathrm{CO}}^{\mathrm{top},-}=1.34\times10^{-7}$ and an upper value of $y_{\mathrm{CO}}^{\mathrm{top},+}=2.48\times10^{-7}$, close to the upper limit of the observational constraint ($y_{\mathrm{CO}}^{\mathrm{top}}=2.0 \pm 0.5\times10^{-7}$). 
Similarly to the Uranus case, results with V20 and V24 both agree with the 1$\sigma$ uncertainty of the V20r network and fit within the measurement error. The V20r network tends to destroy CO faster at higher oxygen abundances. This asymmetry slightly biases the UP toward lower $y_{\mathrm{CO}}^{\mathrm{top}}$ values, as shown by the non-central distribution around the nominal value. Thus, when assessing the corresponding kinetic network induced error on the deep O/H ratio, values tend toward the lower bound. We found a $\sim$ 5\% uncertainty on the oxygen abundance at all latitudes. Around the sub-observer latitude ($\phi\sim25^{\circ}$),  we got $\mathrm{O/H} \sim 186_{-19}^{+9}  \times$ protosolar with a constant $K_{zz}$ and $\mathrm{O/H} \sim 267_{-22}^{+10} \times \mathrm{protosolar}$ otherwise, using a fixed $T^{\mathrm{top}}$. 

The O/H ratios and uncertainties from the kinetic network UP and CO measurement error are summarized for several latitudes in Table \ref{Table : O/H ratios}. These values are obtained from our simulations using a latitudinally dependent $K_{zz}$ but fixed $T^{\mathrm{top}}$.
We compare them with other elemental abundances in all four giant planets. They are compared with other known elemental abundances in all four giant planets in Fig. \ref{fig:SS_enrichments}.

\begin{table}[h]
\centering
\caption{Deep protosolar oxygen enrichment retrieved for a few latitudes on Uranus and Neptune using the latitude-dependent $K_{zz}$ formulation of \cite{Wang2015} and a constant $T^{\mathrm{top}}$.}
\label{Table : O/H ratios}
\small
\setlength{\tabcolsep}{4pt}
\begin{tabular}{c | c c c | c c c}
\hline\hline
& \multicolumn{3}{c}{Uranus} 
& \multicolumn{3}{c}{Neptune} \\
\cline{2-4} \cline{5-7}
Latitude [$^\circ$]  & O/H & Measurement & Model  & O/H & Measurement & Model  \\
   & ($\times$protosolar) & uncertainty & uncertainty  & ($\times$protosolar) & uncertainty & uncertainty \\
\hline
0  & 62  & 61 - 63 & 54 - 69 
   & 222 & 204 - 233 & 198 - 232 \\

30 & 137 & 135 - 138 & 123 - 148
   & 286 & 270 - 298 & 264 - 298 \\

60 & 171 & 169 - 173 & 155 - 186
   & 329 & 311 - 340 & 305 - 342 \\

90 & 177 & 176 - 179 & 161 - 191 
   & 342 & 326 - 355 & 319 - 355 \\
\hline
\end{tabular}

\tablefoot{The first column gives the nominal result, obtained by fitting the observed $y_{\mathrm{CO}}^{\mathrm{top}}$. The second column is obtained by fitting the upper and lower limits of the $y_{\mathrm{CO}}^{\mathrm{top}}$ measurement. The third column is obtained by fitting the 1$\sigma$ range of the CO distribution ($y_{\mathrm{CO}}^{\mathrm{top\pm}}$) derived from 1000 Monte Carlo simulations.}
\end{table}

\subsection{Interpretation}
Although our retrievals show a clear latitudinal dependence in the O/H ratio, they likely bracket a single, homogeneous deep oxygen reservoir rather than revealing different bulk compositions at each sampled latitude, as we further discuss in Sect. \ref{subsect:pseudo2D limits}. The CO abundance we adopted was a disk-averaged quantity and may therefore be weighted toward the central observing latitude, where the modeled vertical mixing efficiency is weaker than near the equator, requiring a higher O/H ratio to fit the CO abundance observed in both planets.   
A simpler assumption to make is that the equatorial values may be the most representative of this deep O/H abundance due to the more efficient vertical mixing. This suggest a varying meridional abundance of CO would result from differentiated mixing in latitude. An analysis of the spatial distribution of the upper tropospheric CO has the potential to help us distinguish the effects of intrinsic latitudinal mixing meridional variations from those introduced by disk averaging.

In the rest of this study, we used the deep oxygen abundances inferred from our thermochemical and diffusion model to compute C/O ratios. We compared these ratios against with a PPD model to gain insights into their formation scenarios. We assumed that our ranges of O/H are a good estimate of the bulk values, particularly at the equator. We discuss the factors that may challenge this assumption in Sect. \ref{section:discussion}.

\section{Constraints on the formation of Uranus and Neptune}
\label{section:formation}

In this section, we employed a 1D PPD model that follows the time-dependent transport of dust particles and vapors, enabling comparison with the ice giant elemental enrichments we derived. 
We first summarize the physical framework used to calculate the radial and temporal evolution of the C/O ratio in the PPD. We then compare these profiles with the deep atmospheric C/O intervals inferred for Uranus and Neptune in the previous section.

\subsection{Protoplanetary disk model}
\label{subsect:ppd_model}

We used the 1D axisymmetric, and vertically integrated PPD model described more extensively in \citet{Aguichine2022}, \citet{Schneeberger2023}, and \citet{Mousis2024}. It captures the coupled evolution and transport along the radial distance $r$ of vapors and ices for each species, including vaporization, condensation, and trapping within clathrate hydrates. The transported quantities are represented by surface density profiles $\Sigma(r,t)$.

The evolution of the gas surface density $\Sigma_{\rm g}$ is governed by the following equation:

\begin{center}
  \begin{equation}
        \frac{\partial \Sigma_{\rm g}}{\partial t} = \frac{3}{r} \frac{\partial}{\partial r} \left[
        r^{1/2} \frac{\partial}{\partial r} \left( r^{1/2}\nu\Sigma_{\rm g} \right) \right]
    \label{eq:ppd_gas_evolution}
,\end{equation}  
\end{center}

\noindent where $\nu$ is the kinematic viscosity, computed following \citet{shakura1973}, dependent on the viscosity parameter $\alpha$, the mean molecular mass of the gas, and the midplane temperature. The latter is calculated by accounting for viscous heating and irradiation by the local environment, represented by an ambient temperature of $T_{\rm amb}=10$~K. Together with the gas surface density, the resulting temperature and pressure profiles determine the locations at which the different volatile phases are thermodynamically stable \citep{Schneeberger2023,Mousis2024}.

The evolution of solids is treated using the two-population framework of \citet{Birnstiel2012}. Disk parameters are adopted from \citet{Schneeberger2023}, with a viscosity parameter $\alpha = 10^{-3}$, a total disk mass of $0.1\,M_\odot$, an outer radius of 200~AU, and an initial mass accretion rate of $10^{-7}\,M_\odot\,\mathrm{yr^{-1}}$. We model the disk evolution between $0$ and $1$~Myr.

For every volatile species $i$, the model separately follows its surface density in the vapor, pure-condensate, and clathrate hydrates reservoirs. The radial and temporal evolution of each reservoir is described by an advection-diffusion equation,

\begin{equation}
    \frac{\partial \Sigma_i}{\partial t} + \frac{1}{r} \frac{\partial}{\partial r} \left[ r \left( \Sigma_iv_i - D_i\Sigma_{\rm g} \frac{\partial}{\partial r} \left( \frac{\Sigma_i}{\Sigma_{\rm g}} \right) \right) \right] - \dot{Q}_i = 0
    \label{eq:ppd_species_transport},
\end{equation}

\noindent where $v_i$ and $D_i$ are the radial velocity and diffusion coefficient of the relevant phase, respectively. The term $\dot{Q}_i$ describes transfers between the vapor, pure-condensate, and clathrate hydrates reservoirs and is positive when material is added to the reservoir.

At each time and radial location in the disk, the partial pressure of each volatile species is compared with the equilibrium vapor pressure of its solid phase. Our simulations include solids in the forms of pure condensates and clathrate hydrates. 
The abundance of a given element relative to H$_2$ is computed by summing the contributions of its molecular carriers in the investigated phase. Elemental ratios (e.g., C/O) are then directly derived from these abundances.

Carbon is assumed to be partitioned among CO, CO$_2$, and CH$_4$, with the remaining oxygen forming H$_2$O. The adopted initial molecular ratios in the PPD are CO:CO$_2$:CH$_4$ = 10:4:1 (see \citealt{Schneeberger2023} for details). 
Using the protosolar abundances of Carbon and Oxygen from from \citet{lodders2025solar}, we deduce the following initial abundances: \ce{CO}/\ce{H2}$=2.654\times 10^{-4}$, \ce{CO2}/\ce{H2}$=1.062\times 10^{-4}$, \ce{CH4}/\ce{H2}$=2.654\times 10^{-5}$ and \ce{H2O}/\ce{H2}$=9.382\times 10^{-4}$.

CO$_2$ does not form clathrate hydrates, as it condenses at temperatures higher than its clathrate stability domain. In contrast, CO and CH$_4$ can be trapped in clathrate hydrates at temperatures exceeding those required for the condensation of their pure ices, provided that sufficient crystalline water is available for their incorporation in the PPD \citep{Mousis2009a,Schneeberger2023,Mousis2024}.

\begin{figure}[h]
\centering
\includegraphics[width = 0.7\linewidth]{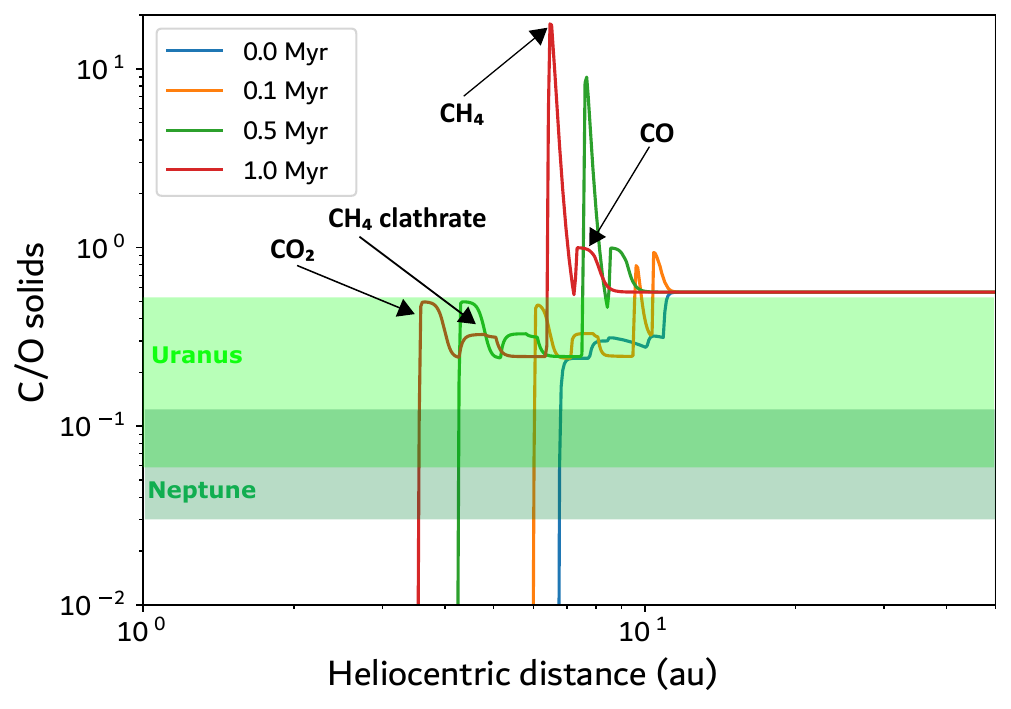}
\caption{Time evolution of the C/O ratio in solids as a function of heliocentric distance in the PPD, including contributions from pure condensates and clathrate hydrates phases. The inflection points trace the ice lines of the dominant C- and O-bearing species (CO$_2$, CH$_4$, CO), as well as the onset of CH$_4$ clathrate hydrates formation, which together structure the radial C/O distribution. 
Arrows indicate the locations of these transitions at 1 Myr. 
The light and dark green shaded regions denote the C/O intervals inferred from the thermochemical modeling. The other shade of green in the middle is the overlap between the ranges of the two planets.}
\label{Fig:Comparison_w_model}
\end{figure}

Figure~\ref{Fig:Comparison_w_model} shows the radial evolution of the C/O ratio in the solid phase (pure condensates and clathrate hydrates) at representative epochs of PPD evolution (0--1 Myr). The inflection points along the curves trace the ice lines of the dominant C- and O-bearing species included in the model. As the disk cools, ice lines progressively migrate inward with time. Each ice line induces a local enrichment in solids, leading to sharp variations in the C/O ratio through the sequestration of volatile carriers from the gas phase. Moving outward from the protosun, the first ice line corresponds to the condensation of \ce{CO2} as a pure ice. This is followed by the formation of \ce{CH4} clathrate hydrates, and at larger distances by the condensation of \ce{CH4} as a pure ice. The outermost ice line corresponds to the condensation of \ce{CO} as a pure ice. CO clathrate hydrates does not form because crystalline water is no longer available in this region. Here, \ce{CO}, the dominant carbon carrier in the outer disk, governs the evolution of the C/O ratio by efficiently removing carbon from the gas phase. Beyond this distance, where all major volatile carriers reside in the solid phase, the C/O ratio converges toward a nearly constant value close to the protosolar value, as illustrated in Fig.~\ref{Fig:Comparison_w_model}.

The radial profile of the C/O ratio in the PPD is highly sensitive to the locations of ice lines and clathration fronts, which are set by the local thermodynamic conditions relative to the equilibrium condensation or clathration curves of each species. 
These locations therefore depend on the adopted molecular abundances and the thermodynamic structure of the disk. The latter is primarily controlled by the disk mass and the mass accretion rate onto the protosun. A more massive disk, or one with a higher accretion rate, undergoes stronger viscous heating, resulting in a warmer PPD and shifting ice lines and clathration fronts to larger heliocentric distances. Because viscous heating dominates the thermal budget, the disk structure is also sensitive to the $\alpha$-viscosity parameter, which regulates the level of turbulence: higher $\alpha$ values lead to higher temperatures throughout the disk.

\subsection{Comparison with the inferred planetary C/O ratios}
\label{subsect:ppd_model_C/O}

Uranus and Neptune probably never reached the pebble isolation mass, meaning that both planets only bind a low-mass envelope that is primarily enriched by volatile species vaporized from icy pebbles \citep{lambrechts2014}. In order to compare the atmospheric C/O constraints with the plausible location and time of formation within the PPD model, we only consider the C/O ratio of the icy pebbles in the PPD model, following the approach of \citet{Mousis2024}. Regions in which the modeled solid-phase (condensates + clathrate hydrates) C/O ratio is in agreement with the atmospheric constraints indicate disk conditions compatible with the inferred deep planetary composition. This requires the assumption that subsequent planetary evolution did not substantially fractionate carbon relative to oxygen, which is out of the scope of this study.

For the planetary O/H ratio, we adopted the atmospheric calculations with a latitude-dependent $K_{zz}$ and a fixed $T^{\rm top}$. We took the corresponding C/H ratio estimated from our thermochemical and diffusion model, which also varies with latitude, as it is linked to the values of $y_\mathrm{CH_4}^{\mathrm{top}}$ and to the oxygen abundance (see eq. 2 and 3 in \citealt{Cavalie2017}). 
On Uranus, we obtained $\mathrm{C/H} \sim [21$--$56] \times \, \mathrm{protosolar}$ while on Neptune we found $\mathrm{C/H} \sim [19$--$47] \times \, \mathrm{protosolar}$, with both ratios decreasing with latitude. We combine these carbon abundances with the retrieved deep oxygen abundances. For Uranus, our thermochemical model constrains $\mathrm{C/O} \sim [0.06$--$0.52]$ while for Neptune $\mathrm{C/O} \sim [0.02$--$0.12]$, also declining from the equator to the pole.

The PPD model predicts a protosolar C/O ratio at the present-day orbital distances of Uranus and Neptune ($\sim$20--30 AU).
In the Uranus case, the high-end of the C/O range inferred from disequilibrium species is near protosolar. The highest C/O ratios correspond to low latitudes. This is broadly consistent with the PPD model, as the present-day C/O falls close to the range predicted at its current orbital distance. In contrast, this approach fails to explain the subsolar C/O ratio inferred for Neptune.
Some C/O values, however, are found to be plausible in both planets, that is, $\mathrm{C/O} \sim [0.06$--$0.12]$. This corresponds to the value found from mid-to-high latitudes for Uranus and for $\phi<30^{\circ}$ for Neptune. Enrichment on both planets might be reconciled if the C/O ratio in Uranus tends toward the lower end of the range, which could be explained if $K_{zz}$ is lower on Uranus than on Neptune. This is further discussed in Sect. \ref{subsect:uncertainty kzz}.

A subsolar C/O ratio in Uranus emerges naturally in our model if its building blocks formed interior to the \ce{CH4} ice line, where methane is largely depleted from the solid phase. The subsolar C/O ratio of Neptune can be explained if its building blocks formed near the \ce{H2O} ice line. In this region, solids are dominated by water ice, while carbon remains largely in the vapor phase (\ce{CO2}, \ce{CH4}, CO), thereby limiting the efficiency of carbon incorporation into the growing planet. A fraction of carbon may also have been sequestered in refractory organics and rocks, but only to a limited extent to remain consistent with the inferred subsolar C/O ratio. This interpretation disfavors in situ formation for Neptune and instead supports formation closer to the Sun followed by outward migration (e.g., \citealt{Raymond2022,Mousis2018,Mousis2020,helled2020uranus}).

\section{Discussion}

\label{section:discussion}
We used a pseudo-2D thermochemical and diffusion model to provide a reliable range of the deep oxygen abundance using upper tropospheric measurements of the CO mole fraction on Uranus and Neptune. We accounted for the meridional variation of several key parameters to derive a range of plausible values. For the first time, we translated the propagation of reaction rate uncertainties into deep oxygen abundance retrievals. We found that the meridional variability in vertical mixing dominates the inferred deep oxygen abundance. Here we discuss the various aspects of the existing model that could be further improved, and the main uncertainties that affect our results.

\begin{figure}[h]
        \centering
        \includegraphics[width = 0.7\linewidth]{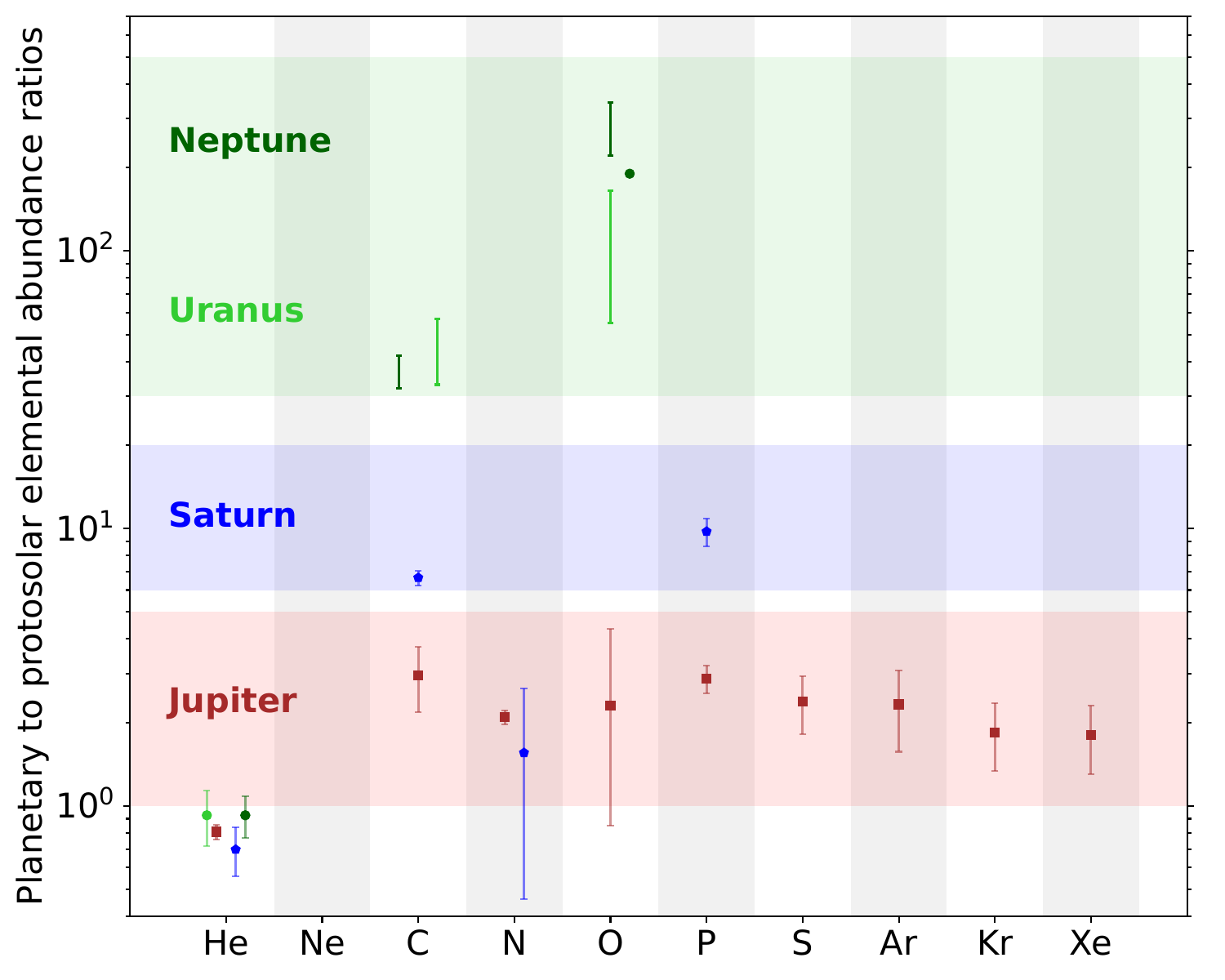}
        \caption{Elemental abundances for all four giant planets of the Solar System, expressed with respect to the protosolar values of \cite{lodders2025solar} and the observational data from Table \ref{table:compo_SS}. The oxygen intervals for Uranus and Neptune correspond to the latitudinal range of values obtained with a varying $K_{zz}$ and a fixed $T^{\mathrm{top}}$. The single oxygen point represents the result of \cite{Venot2020} for Neptune.}
        \label{fig:SS_enrichments}
\end{figure}

\subsection{Uncertainties on $K_{zz}$} \label{subsect:uncertainty kzz}
The largest uncertainty in our model arises from the magnitude of the eddy diffusion coefficient. The MLT formulation predicts a value of $10^8$ cm\textsuperscript{2}\,s\textsuperscript{-1}, usually taken as a constant throughout the troposphere of the ice giants \citep{Cavalie2017}. When accounting for planetary rotation, following the approach of \citet{Wang2015}, the $K_{zz}$ value drops by several orders of magnitude near the quench level. This results in a higher oxygen abundance being required at the poles to fit the observed CO value. Since this formula suggests that atmosphere is better mixed near the equator, we can expect that the lower latitudes are more representative of the bulk abundance of oxygen. The prescription of \cite{Wang2015} also has 10\% and 20\% uncertainties on the low latitude and high latitude scalings, respectively (see eq. \ref{eq:Kzz_lowlat} and \ref{eq:Kzz_highlat}). We applied this latitudinally dependent $K_{zz}$ in the entire troposphere, as illustrated by Fig. \ref{fig:Kzz_both}, but it is only valid in the convective region. If convection is inhibited at the water-cloud base, the value of $K_{zz}$ could be as low as molecular diffusivity, that is, on the order of $10^{-3}$ cm\textsuperscript{2}\,s\textsuperscript{-1} \citep{Cavalie2017}. 

This local $K_{zz}$ reduction due to convection inhibition is not limited to the water cloud. At low-to-mid latitudes on Uranus and all latitudes on Neptune, methane is abundant enough that convection inhibition can occur \citep{clement2024}. Using a cloud resolving model, these authors showed that this can also lead to $K_{zz} \sim10^{-3}$ cm\textsuperscript{2}\,s\textsuperscript{-1}. Meanwhile, \cite{Ge2024} showed analytically that when taking into consideration a more realistic hydrological cycle involving precipitation, the upper limit of $K_{zz}$ due to moist convection in the upper troposphere is multiple orders of magnitude lower than that predicted by MLT and by the formalism we used. The exact magnitude of $K_{zz}$ in the water condensation region remains uncertain. 

Using 2D hydrodynamic simulations, \cite{Yang2026} showed that it is possible to derive a more self-consistent $K_{zz}$ profile on Jupiter, by inverting their quasi steady-state tracer transport equation. With similar hydrodynamic simulations, \cite{hyder2025supersolar} first studied the effect of trace species transport across the lifting condensation level (LCL) by short-term dynamics. \cite{hyder2025supersolar} showed that while the CO abundance might be affected across the water-cloud level over short timescales, simulations run over longer timescales produce a stabilization of the CO vertical profile, which is consistent with a coupled chemical kinetic transport model \citep{Yang2026}.

Interestingly, according to \cite{ge2025}, falling precipitation beneath the water-cloud LCL on Jupiter causes a large-scale drying of the atmosphere at the mid-latitudes. The authors show that the meridional depletion of water at these levels (and the corresponding increase in stratification) can be explained by the effect of planetary rotation rather than cloud microphysics. Similar effects due to planetary rotation are expected to occur in Uranus and Neptune, which can help constrain the deep meridional structure of both planets, in combination with spatially resolved CO observations.

Compared to these hydrodynamic simulations, our thermochemical model offers a comprehensive description of chemical processes, though with simplified transport processes. In theory, it can be coupled with cloud microphysics models, which would be especially important to model the sulfur chemical cycle. Recent constraints on the opacities, aerosol composition and size distribution (\citealt{Irwin2022,Toledo2025}), or cloud density \citep{Ge2024} in the upper tropospheric main cloud deck will help in that regard. 

Some evidence indicates a different deep convective activity between the ice giants. The magnitude difference of almost two orders in the CO abundance between the two planets and the order-of-magnitude difference in internal heat flux release (see Table \ref{table:constants planets}), might indeed translate into a lower deep $K_{zz}$ on Uranus. 

Instead, when we assume that the O/H and C/O ratios are similar in both planets, we can estimate the deep $K_{zz}$ ratio between Uranus and Neptune. We can take the O/H derived at the equator with $K_{zz} = 10^8$ cm$^2$s$^{-1}$ on Neptune as a reference for both planets ($\mathrm{O/H}\sim$ 180 $\times \mathrm{protosolar}$). Then, we computed the deep $K_{zz}$ value that fits the measured $y_{CO}^{top}$ on Uranus at the equator, using $\mathrm{O/H} \sim$ 180 $\times$ protosolar as a reference. In this scenario, we found that around the quench level, Uranus' $K_{zz}$ is smaller than Neptune's by a factor of $\sim$ 5000. 
However, at such pressures ($ p\gtrsim  1000$ bars), $K_{zz}$ might also be similar on both planets. Instead, $K_{zz}$ could only be decreased at lower pressure levels, that is, around the water-cloud base due to convection inhibition. In this case, no direct conclusion on the deep $K_{zz}$ difference between the two planets can be drawn from the quenched CO abundance and the C/O ratio solely.

The true magnitude of $K_{zz}$ remains loosely constrained, and may be affected by rotation, magnetic fields \citep{Stevenson1979} and convection inhibition due to condensation \citep{clement2024}. New experimental work could be used to improve the current formulation (e.g., \citealp{Lepot2018, Bouillaut2021}).
An atmospheric probe as part of the UOP mission could favor an entry site around the equator for an increased chance of probing a CO abundance more indicative of the bulk value. Such a probe might also provide constraints on $K_{zz}$ through measurements of the vertical profile of additional trace species \citep{Atreya1999,Wong2004}.

\subsection{Completeness of chemical network and cloud model}

Our thermochemical model relies on continuously evolving chemical networks. As wider pressure and temperature ranges can be explored experimentally and more kinetic data are acquired, larger and more accurate networks are developed. We used the reduced version of the V20 network to derive the deep O/H ratio on both planets. We compared the CO profiles obtained at a similar O/H ratio for two more complete C/H/O/N networks, and found that they both reproduce the result of V20r well within the complete network uncertainty range. This is particularly useful because the time required to solve the continuity equation (eq. \ref{eq:continuity equation}) increases nonlinearly with the size of the chemical scheme. Thus, for computational efficiency reasons, we chose to proceed with the V20r network. The relative error on the measured CO abundance is much lower in the Uranus case than in the Neptune case. We showed that while the error on the CO abundance retrieval due to the UP of the kinetic network are on the same order as the CO measurement uncertainty on Neptune, it is much larger in the Uranus case. This highlights the importance of higher accuracy on the parametrized reaction rates.

Detection of \ce{H2S} in the near-infrared above the cloud in the upper troposphere of both planets (\citealt{Irwin2018,Irwin2019}) highlights the importance of accounting for sulfur in chemical networks. The sulfur-nitrogen coupling is particularly important due to the expected presence of \ce{NH4SH} and \ce{H2S} clouds in the troposphere, and their properties depend on the \ce{H2S} and \ce{NH3} abundances \citep{Hueso2019}. \cite{veillet2026} recently updated the V24 network and added sulfur species, which made the first C/H/O/N/S chemical scheme applicable to the troposphere of giant planets with experimental validation. Several disequilibrium sulfur species are thought to be present (e.g., \ce{CH3SH}), but none have been observed in the upper troposphere yet \citep{Moses2020}. 

A cloud resolving model would have to be coupled with such a thermochemical model to better constrain the deep sulfur abundance. The implementation of a condensation scheme for sulfur species in our model can be complex if \ce{H2S} is abundant enough to locally inhibit convection. The latitudinal variation of the abundance of \ce{H2S} inferred below the cloud could further constrain our model \citep{Molter2021, tollefson2019}. While this development would help us to infer the deep S/H ratio but not directly constrain the deep O/H ratio, it is a useful consideration to compare the elemental abundances together to differentiate the formation scenarios of giant planets  \citep{Mousis2018}. 

Phosphorus-bearing species are the next priority for chemical networks, as \ce{PH3} is expected to be quenched and has an upper limit in the upper troposphere of both planets \citep{encrenaz1996b,Teanby2019}. Similar to \ce{CO}, \ce{PH3} is in thermochemical equilibrium with water in the deep troposphere. It could serve as a proxy to further constrain the magnitude of $K_{zz}$ around the quench level (see \citealt{wang2016,Visscher2005,Visscher2006} on Jupiter), and thus to indirectly constrain the deep oxygen abundance.

Another approach to build a chemical network is to use a reaction mechanism generator (RMG) method. This allows us to build a chemical scheme mathematically, based on graphs and trees \citep{Gao2016}. This has recently been applied to model the CO abundance in the troposphere of Jupiter \citep{Yang2026}. Their network included the key reaction \ce{H + CH3OH \rightarrow CH3 + H2O} from \cite{Hidaka1989}, which was present in the V12 network but was removed in the V20 network. With their RMG network implemented in a 1D thermochemical and diffusion model, and applying a constant eddy diffusion coefficient profile fixed at $K_{zz} = 10^8$ cm$^2$ s$^{-1}$, \cite{Yang2026} found a subsolar deep oxygen enrichment  $\mathrm{O/H} = [0.5-0.6] \times \odot$ on Jupiter. With the same $K_{zz}$ profile and instead using the V20 network, \cite{Cavalie2023} also derived a subsolar enrichment of $\mathrm{O/H} = 0.3 \times \odot$. 
However, \citet{Yang2026} also coupled their RMG network with a 2D hydrodynamic and cloud-resolving framework. The simulations of \citet{Yang2026} support a modestly supersolar Jovian oxygen enrichment of $\mathrm{O/H} \sim [1.0-1.5] \times \odot$, with the long-term evolution of the horizontally averaged CO profiles favoring values near the upper part of this range.
No such RMG-based network has been applied to the troposphere of the ice giant planets yet, and future work could include comparing a RMG-generated chemical network with those validated against experimental data acquired in combustion chambers, such as the one we used here.

\subsection{Temperature structure}

The vertical thermal structure is also subject to ambiguity. \cite{Cavalie2017} showed that moist convection inhibition at the water-cloud level can strongly affect the retrieved deep oxygen abundance on the ice giant planets. We used similar three-layer thermal profiles, producing a temperature jump at the water-cloud level that results in  deep tropospheric temperature hundreds of Kelvin warmer when compared to using a classical dry or wet adiabat.  We start extrapolating our temperature profile from below the methane cloud to avoid any convection inhibition or superadiabaticity effects as prescribed by \cite{Cavalie2017}. In this three-layer formulation the atmosphere is fully saturated, all condensates instantly precipitate, and no latent heat release is accounted for \citep{Guillot1995}. However, \cite{ge2023moist} showed that the assumption of a fully saturated medium may not always be verified. In this case, the author shows that the criterion for convection inhibition may require a lower abundance threshold of the condensate than the prescription of \cite{Leconte2017}. This could theoretically be applied to the \ce{H2S}, \ce{NH3}, and \ce{NH4SH} cloud deck.

To compute the radiative temperature gradient, we used a Rosseland mean opacity parametrized for a varying metallicity, as prescribed by \cite{Valencia2013}. This prescription is based on the Rosseland opacity tables of \cite{Freedman2008}, computed for a broad range of pressure and temperature over a discrete set of metallicities. As emphasized by \cite{Cavalie2017}, these are based on older opacity data, do not account for the opacity of the condensate and are poorly constrained for heterogeneous compositions. Recently, \cite{siebenaler2025b} published the first cloudy mean opacity tables while also using the most recent molecular cross section data. These could potentially be adapted to model the higher metallicities in the troposphere of the ice giants.

\subsection{Pseudo-2D limitations}
\label{subsect:pseudo2D limits}
With our pseudo 2D model, we studied the effect of varying the upper tropospheric temperature and methane mole fraction together with latitude. These variations can positively or negatively interfere depending on the meridional structure of $T^{\mathrm{top}}$, causing a temperature gradient between the equator and the poles. However, the radiative-convective model yields much stronger temperature variations than those observed in Uranus and Neptune \citep{milcareck2024}. In contrast, accounting for latitudinal variations in upper troposphere temperature has a much smaller effect on the inferred O/H ratio than including a latitudinally varying $K_{zz}$ compared with simulations in which these parameters are held constant.

Extrapolating thermal profiles in such a pseudo-2D manner is moreover challenging because the horizontal structure of the deep troposphere, namely, below the main cloud deck, might be homogeneous with latitude. On Uranus, the \ce{CH4} abundance may simply be constant with latitude at pressure greater than $\sim$5 bars according to \cite{Sromovsky2019}. These authors found that this apparent latitudinal uniformity of the methane mole fraction at depth does not strongly depend on whether a step or descended profile is used, on which the methane vertical profile is parametrized to smoothly drop down with increasing pressure levels. The same conclusion is found about the methane reservoir on Neptune \citep{tollefson2019,Irwin2021}. Moreover, a methane mole fraction latitudinal variation would induce strong humidity winds due to the mean molecular weight gradient, which are not observed as stated by \cite{tollefson2018}. Ultimately, the \ce{CH4} abundance variation with latitude that is observed on both planets may not extend deeper than a few bars, and would then be the result of condensation and upper tropospheric circulation patterns. \ce{H2S} is also expected to be well-mixed below the \ce{NH4SH} cloud, located in the 30--50 bars region \citep{Molter2021,tollefson2019}. 
We therefore could have computed our C/O ratios with a latitudinally independent deep carbon abundance inferred from observations, namely $\mathrm{C/H} \sim [33$--$57] \times \, \mathrm{protosolar}$ for Uranus \citep{Sromovsky2019} and  $\mathrm{C/H} \sim [32$--$42] \times \, \mathrm{protosolar}$ for Neptune \citep{tollefson2019}. However, because our modeled latitudinally dependent C/H ratios agree with the aforementioned observational constraints, they were chosen to compute a more self-consistent C/O ratio for both planets.

Additionally, our approach revolves around fitting the same $y_{\mathrm{CO}}^{\mathrm{top}}$ value at all latitudes. However, if we assume a homogeneous deep oxygen reservoir at all latitudes while expecting planetary rotation to play a role in the horizontal distribution of disequilibrium species, we would need spatially resolved latitude CO measurements. On Jupiter, such nonuniform latitudinal abundances have been observed for other quenched species like \ce{GeH4} and \ce{PH3} \citep{grassi2020}. A future approach could then be to constrain the $K_{zz}$ meridional variation based on a given $y_{\mathrm{CO}}^{\mathrm{top}}$ latitudinal profile. 

On Uranus, the fitted $y_{CO}^{top}$ corresponds to observations at the central meridian and $\phi= 60^{\circ}$. At this latitude, around the quench level, the formulation of \cite{Wang2015} predicts $K_{zz}\sim10^5$ cm\textsuperscript{2}\,s\textsuperscript{-1}, that is, lower by three orders of magnitude than the equatorial and MLT values. 
At this latitude, with fixed $T^{\mathrm{top}}$ and constant $K_{zz}$, we found $\mathrm{O/H} \sim 50_{-6}^{+5}  \times \mathrm{protosolar}$. Meanwhile, our model predicts $\mathrm{O/H} \sim 158_{-15}^{+12} \times \mathrm{protosolar}$, using the latitude-dependent $K_{zz}$ profile. These two values are thus incompatible by a factor of 3.
Similarly, on Neptune, at $\phi= 25^{\circ}$, our model predicts $\mathrm{O/H} \sim 186_{-19}^{+9} \times \mathrm{protosolar}$, and $\mathrm{O/H} \sim 267_{-22}^{+10} \times \mathrm{protosolar}$ with constant and latitude-dependent $K_{zz}$, respectively. 
This difference shows that convection inhibition due to planetary rotation can be a key process to infer the deep elemental abundances.
Therefore, the derived enrichments at non-equatorial latitudes might not be representative of the bulk oxygen abundance. Future in situ measurements by a mission such as UOP, particularly at low latitudes where vertical mixing is most efficient, will be essential to test these predictions and to refine the chemical networks and the treatment of moist convection in the ice giants' atmospheres.

\subsection{Limitations of the formation models}
\label{ss:disk_model_limitation}

The disk evolution model we used to compute the primordial composition of the PSN also relies on several assumptions. One of them concerns the initial volatile abundances considered: as described in Sect. \ref{section:formation}, the elements are allocated to molecules from their protosolar abundance described in \citet{lodders2025solar}. Thus the protosolar carbon abundance is split between the main carbon bearing molecules (CO, CO$_2$, CH$_4$) according to the ratio found in 67P \citep{Mousis2014,Leroy2015}, following the approach of \cite{Schneeberger2023}. However, the presence of organic matter, made of carbon, should also influence its distribution between molecules, and some studies have considered that half the primordial carbon was in refractory molecules \citep{Bergin_Blake_Ciesla_Hirschmann_Li_2015,Mousis2024}. Similarly, other less abundant C-bearing molecules, such as methanol, were not considered in the computing of the composition of the PSN.

Another assumption comes from the clathrate hydrate formation model. The clathrate hydrate dissociation pressure equations we used were taken from \citet{Lunine1985}, \citet{Hersant_Gautier_Lunine_2004}, and \citet{Longhi_2005}, who extrapolated them from experimental measurements made at different pressures and temperatures than that of the disk.

Finally, the initial parameters of the disk evolution model are set based on values commonly used in the community. First, the $\alpha$-viscous parameter is set to 10$^{-3}$, to be consistent with values usually taken between $10^{-4}$ and  $10^{-2}$ \citep{Hartmann1998, Hueso_Guillot_2005, Desch2017}. Then, the initial mass accretion rate is set to $\dot{M}_{\mathrm{acc},0} = 5 \times 10^{-7}\,M_\odot\,\text{yr}^{-1}$, based on observations of values between $10^{-9}$ and $10^{-6}\,M_\odot\,\text{yr}^{-1}$ in PPDs \citep{Hartmann1998, Gullbring1998}. 
These variations will impact the positions of the clathration lines and ice lines, as well as the magnitude of the enrichment. On the one hand, the higher the accretion rate, the warmer the disk, and therefore, the farther out the lines. On the other hand, when we increased the  $\alpha$-viscous parameter, the enrichments at the lines locations were lower, but the peaks were broader.  In our case, this would decrease the C/O ratio at the carbon-bearing species lines and increase the ratio at the water-ice line.

Furthermore, there is an uncertainty regarding the formation of Uranus and Neptune. Considering their large differences in size, composition, and internal structures compared to the gas giants Jupiter and Saturn, it is highly probable that their formation pathway was different. \cite{lambrechts2014} proposed that their formation could have stopped at the pebble accretion stage, as they could not reach the pebble isolation mass, which is a critical condition for initiating a runaway gas accretion phase. In that scenario, Uranus and Neptune could only have accreted volatiles from solids and only a small fraction of vapor from the PPD, which explains their high metallicity. 

\section{Conclusion} \label{section:conclusion}
We reassessed the enrichment in protosolar oxygen on Uranus and Neptune with a pseudo-2D thermochemical diffusion framework. We accounted for meridional variations in temperature structure, methane abundance, and vertical mixing from observations and models. With CO now being unambiguously detected in the troposphere of both planets, we matched the deep O/H ratio as a function of latitude.
We thus derived an interval of possible deep O/H ratios. Based on the higher mixing efficiency at low latitudes, we suggest that the equatorial values might provide a better proxy for the bulk composition. We also assessed the impact of the uncertainty propagation in a chemical network and on the aforementioned O/H ratio. A range of C/O ratios was subsequently computed for both planets and was tested against a protoplanetary disk evolution model. 
Our findings support the need for in situ measurements, namely with the UOP mission, and for further constraints on the magnitude of $K_{zz}$ in the troposphere of ice giant planets, accounting for planetary rotation and ice condensation. \\

\noindent The main conclusions of this work are listed below.

\begin{enumerate}
    \item Our pseudo‑2D thermochemical modeling showed that latitudinal variations in vertical mixing dominate the inferred deep oxygen enrichment of Uranus and Neptune, while the uncertainties in the chemical kinetics play a secondary but quantifiable role.\\
    \item On Uranus, using a constant $T^{\mathrm{top}}$ and a varying $K_{zz}$ profile with latitude, we found a deep oxygen abundance ranging from $62 - 177 \times \mathrm{ protosolar}$ and $54 - 191 \times \mathrm{ protosolar}$ when considering the chemical network uncertainties. The equatorial value is $\mathrm{O/H} \sim 62_{-8}^{+7} \times \mathrm{protosolar}$. At the central  measurement latitude $\phi = 60^{\circ}$, we found $\mathrm{O/H} \sim 171_{-16}^{+15} \times \mathrm{protosolar}$. \\
    \item With the same considerations, we found a deep oxygen abundance on Neptune, ranging from $222 - 342 \times \mathrm{protosolar}$ and $199 - 355 \times \mathrm{protosolar}$ when considering the chemical network uncertainties. The equatorial value is  $\mathrm{O/H} \sim 222_{-18}^{+11} \times \mathrm{protosolar}$. At the central measurement latitude $\phi = 25^{\circ}$, we found $\mathrm{O/H} \sim 267_{-22}^{+10} \times \mathrm{protosolar}$.\\
    \item The reduced chemical network V20r reproduces the observed upper tropospheric CO abundance on Neptune within its propagated $1\sigma$ uncertainty range. On Uranus, the uncertainty from chemical kinetics exceeds the observational uncertainty on CO. In addition, the results from V20r are consistent with those obtained using the more complete C/H/O/N chemical networks on both planets. \\
    \item The resulting largely supersolar and different O/H values between the two planets result in a C/O ratio that is close to protosolar for Uranus and subsolar for Neptune. If the C/O ratio derived at low latitudes by our thermochemical modeling is most representative of the bulk value and the vertical mixing in both planets scales similarly (see Sect. \ref{subsect:kzz}), our results suggest a different formation pathway for Uranus and Neptune. \\
\end{enumerate}

\clearpage
\section*{Acknowledgements}
      V. Hue and T. Briand acknowledge support from the French government under the France 2030 investment plan, as part of the Initiative d’Excellence d’Aix-Marseille Université – A*MIDEX AMX-22-CPJ-04. V. Hue, T. Briand and T. Cavalié acknowledge support from CNES. T. Briand thanks G. Milcareck for providing his radiative-convective 2D temperature structure from his 2024 paper. T. Briand, and V. Hue thank P. Irwin for providing the methane mole fraction profiles from his 2021 paper. T. Briand and V. Hue thank O. Venot and M. Roman for insightful discussions about the core concepts of the present work. We thank A. Hyder for his thorough review of this paper.

\bibliographystyle{abbrvnat} 
\bibliography{biblio}

@ARTICLE{Aguichine2022,
       author = {{Aguichine}, Artyom and {Mousis}, Olivier and {Lunine}, Jonathan I.},
        title = "{The Possible Formation of Jupiter from Supersolar Gas}",
      journal = {Planet. Sci. J.},
         year = 2022,
       volume = {3},
       number = {6},
          eid = {141},
        pages = {141},
archivePrefix = {arXiv},
       eprint = {2204.14102},
 primaryClass = {astro-ph.EP},
       adsurl = {https://ui.adsabs.harvard.edu/abs/2022PSJ.....3..141A}
}

@article{Fletcher2020,
  author  = {Fletcher, Leigh N. and de Pater, Imke and Orton, Glenn S.
             and Hofstadter, Mark D. and Irwin, Patrick G. J.
             and Roman, Michael T. and Toledo, Daniel},
  title   = {Ice Giant Circulation Patterns: Implications for Atmospheric Probes},
  journal = {Space Sci. Rev.},
  year    = {2020},
  volume  = {216},
  number  = {2},
  pages   = {21},
}

@article{shakura1973,
  title={Black holes in binary systems. Observational appearance.},
  author={Shakura, Nicolai Ivanovich and Sunyaev, Rashid Alievich},
  journal={A\&A},
  volume={24},
  pages={337--355},
  year={1973}
}

@article{Hofstadter2024,
  title={Uranus Study Report: KISS},
  author={Hofstadter, Mark and Helled, Ravit and Stevenson, David J and Ehlmann, Bethany and Bethkenhagen, Mandy and Cao, Hao and Dong, Junjie and El Moutamid, Maryame and Ermakov, Anton and Fuller, Jim and others},
  journal={arXiv e-prints},
  pages={arXiv--2412},
  year={2024}
}

@article{Cavalie2026,
  title={Detection of stratospheric HCN and tropospheric CO in Uranus and the implication for their sources},
  author={Cavali{\'e}, T and Moreno, R and Lefour, C and Luszcz-Cook, SH and Fouchet, T and Lellouch, E and Carri{\'o}n-Gonz{\'a}lez, O and Benmahi, B and Guerlet, S and Milcareck, G and others},
  journal={A\&A},
  volume={710},
  pages={A362},
  year={2026},
  publisher={EDP Sciences}
}

@article{Lepot2018,
  title={Radiative heating achieves the ultimate regime of thermal convection},
  author={Lepot, Simon and Auma{\^\i}tre, S{\'e}bastien and Gallet, Basile},
  journal={Proc. Natl. Acad. Sci.},
  volume={115},
  number={36},
  pages={8937--8941},
  year={2018},
  publisher={National Academy of Sciences}
}

@article{Bouillaut2021,
  title={Experimental observation of the geostrophic turbulence regime of rapidly rotating convection},
  author={Bouillaut, Vincent and Miquel, Benjamin and Julien, Keith and Auma{\^\i}tre, S{\'e}bastien and Gallet, Basile},
  journal={Proc. Natl. Acad. Sci.},
  volume={118},
  number={44},
  pages={e2105015118},
  year={2021},
  publisher={National Academy of Sciences}
}

@article{Hueso_Guillot_2005, title={Evolution of protoplanetary disks: constraints from DM Tauri and GM Aurigae}, volume={442}, number={2}, journal={A\&A}, author={Hueso, R. and Guillot, T.}, year={2005}, pages={703–725}, language={en} }

@article{Longhi_2005, title={Phase equilibria in the system CO 2-H 2O I: New equilibrium relations at low temperatures}, volume={69}, journal={Geochimica et Cosmochimica Acta}, author={Longhi, John}, year={2005}, pages={529–539} }

@article{Hersant_Gautier_Lunine_2004, title={Enrichment in volatiles in the giant planets of the Solar System}, volume={52}, journal={Planet. Space Sci.}, author={Hersant, F. and Gautier, D. and Lunine, J. I.}, year={2004}, pages={623–641} }

@article{Bergin_Blake_Ciesla_Hirschmann_Li_2015, title={Tracing the ingredients for a habitable earth from interstellar space through planet formation}, volume={112}, journal={Proc. Natl. Acad. Sci.}, author={Bergin, Edwin A. and Blake, Geoffrey A. and Ciesla, Fred and Hirschmann, Marc M. and Li, Jie}, year={2015}, pages={8965–8970} }

@ARTICLE{Gullbring1998,
       author = {{Gullbring}, Erik and {Hartmann}, Lee and {Brice{\~n}o}, Cesar and {Calvet}, Nuria},
        title = "{Disk Accretion Rates for T Tauri Stars}",
      journal = {ApJ},
         year = 1998,
       volume = {492},
       number = {1},
        pages = {323-341},
       adsurl = {https://ui.adsabs.harvard.edu/abs/1998ApJ...492..323G}
}

@ARTICLE{Desch2017,
       author = {{Desch}, Steven J. and {Estrada}, Paul R. and {Kalyaan}, Anusha and {Cuzzi}, Jeffrey N.},
        title = "{Formulas for Radial Transport in Protoplanetary Disks}",
      journal = {ApJ},
         year = 2017,
       volume = {840},
       number = {2},
          eid = {86},
        pages = {86},
archivePrefix = {arXiv},
       eprint = {1704.01267},
 primaryClass = {astro-ph.EP},
       adsurl = {https://ui.adsabs.harvard.edu/abs/2017ApJ...840...86D}
}

@ARTICLE{Hartmann1998,
       author = {{Hartmann}, Lee and {Calvet}, Nuria and {Gullbring}, Erik and {D'Alessio}, Paola},
        title = "{Accretion and the Evolution of T Tauri Disks}",
      journal = {ApJ},
         year = 1998,
       volume = {495},
       number = {1},
        pages = {385-400},
       adsurl = {https://ui.adsabs.harvard.edu/abs/1998ApJ...495..385H}
}

@article{lambrechts2014,
  title = {Separating Gas-Giant and Ice-Giant Planets by Halting Pebble Accretion},
  author = {Lambrechts, M. and Johansen, A. and Morbidelli, A.},
  year = 2014,
  journal = {A\&A},
  volume = {572},
  pages = {A35},
  urldate = {2024-02-07}
}

@article{Atreya1999,
  title={A comparison of the atmospheres of Jupiter and Saturn: deep atmospheric composition, cloud structure, vertical mixing, and origin},
  author={Atreya, SK and Wong, MH and Owen, TC and Mahaffy, PR and Niemann, HB and De Pater, I and Drossart, P and Encrenaz, Th},
  journal={Planet. Space Sci.},
  volume={47},
  number={10-11},
  pages={1243--1262},
  year={1999},
  publisher={Elsevier}
}

@article{Stevenson1979,
  title={Turbulent thermal convection in the presence of rotation and a magnetic field: a heuristic theory},
  author={Stevenson, David J},
  journal={Geophys. Astrophys. Fluid Dyn.},
  volume={12},
  number={1},
  pages={139--169},
  year={1979},
  publisher={Taylor \& Francis}
}

@article{Irwin2016,
  title={Time variability of Neptune’s horizontal and vertical cloud structure revealed by VLT/SINFONI and Gemini/NIFS from 2009 to 2013},
  author={Irwin, PGJ and Fletcher, Leigh Nicholas and Tice, D and Owen, SJ and Orton, GS and Teanby, NA and Davis, GR},
  journal={Icarus},
  volume={271},
  pages={418--437},
  year={2016},
  publisher={Elsevier}
}

@article{Carrion2023,
  title={Doppler wind measurements in Neptune’s stratosphere with ALMA},
  author={Carri{\'o}n-Gonz{\'a}lez, {\'O}scar and Moreno, Raphael and Lellouch, Emmanuel and Cavali{\'e}, Thibault and Guerlet, Sandrine and Milcareck, Gwena{\"e}l and Spiga, Aymeric and Cl{\'e}ment, No{\'e} and Leconte, J{\'e}r{\'e}my},
  journal={A\&A},
  volume={674},
  pages={L3},
  year={2023},
  publisher={EDP Sciences}
}

@article{Cavalie2024,
  title={The deep oxygen abundance in solar system giant planets, with a new derivation for Saturn},
  author={Cavali{\'e}, Thibault and Lunine, Jonathan and Mousis, Olivier and Hueso, Ricardo},
  journal={Space Sci. Rev.},
  volume={220},
  number={1},
  pages={8},
  year={2024},
  publisher={Springer}
}

@article{Fletcher2011,
  title={Thermal structure and dynamics of Saturn’s northern springtime disturbance},
  author={Fletcher, Leigh N and Hesman, Brigette E and Irwin, Patrick GJ and Baines, Kevin H and Momary, Thomas W and Sanchez-Lavega, Agustin and Flasar, F Michael and Read, Peter L and Orton, Glenn S and Simon-Miller, Amy and others},
  journal={Science},
  volume={332},
  number={6036},
  pages={1413--1417},
  year={2011},
  publisher={American Association for the Advancement of Science}
}

@article{Wong2004,
  title={Updated Galileo probe mass spectrometer measurements of carbon, oxygen, nitrogen, and sulfur on Jupiter},
  author={Wong, Michael H and Mahaffy, Paul R and Atreya, Sushil K and Niemann, Hasso B and Owen, Tobias C},
  journal={Icarus},
  volume={171},
  number={1},
  pages={153--170},
  year={2004},
  publisher={Elsevier}
}

@article{Mahaffy2000,
  title={Noble gas abundance and isotope ratios in the atmosphere of Jupiter from the Galileo Probe Mass Spectrometer},
  author={Mahaffy, PR and Niemann, HB and Alpert, A and Atreya, SK and Demick, J and Donahue, TM and Harpold, DN and Owen, TC},
  journal={J. Geophys. Res.},
  volume={105},
  number={E6},
  pages={15061--15071},
  year={2000},
  publisher={Wiley Online Library}
}

@article{Encrenaz2004,
  title={First detection of CO in Uranus},
  author={Encrenaz, Th and Lellouch, Emmanuel and Drossart, Pierre and Feuchtgruber, Helmut and Orton, Glenn S and Atreya, Sushil K},
  journal={A\&A},
  volume={413},
  number={2},
  pages={L5--L9},
  year={2004},
  publisher={EDP Sciences}
}

@article{Visscher2006,
  title={Atmospheric chemistry in giant planets, brown dwarfs, and low-mass dwarf stars. II. Sulfur and phosphorus},
  author={Visscher, Channon and Lodders, Katharina and Fegley, Jr, Bruce},
  journal={ApJ},
  volume={648},
  number={2},
  pages={1181--1195},
  year={2006}
}

@article{Visscher2005,
  title={Chemical constraints on the water and total oxygen abundances in the deep atmosphere of Saturn},
  author={Visscher, Channon and Fegley, Jr, Bruce},
  journal={ApJ},
  volume={623},
  number={2},
  pages={1221--1227},
  year={2005}
}

@article{Marten1993,
  title={First observations of CO and HCN on Neptune and Uranus at millimeter wavelengths and the implications for atmospheric chemistry},
  author={Marten, A and Gautier, D and Owen, T and Sanders, DB and Matthews, HE and Atreya, SK and Tilanus, RPJ and Deane, JR},
  journal={ApJ},
  volume={406},
  pages={285--297},
  year={1993}
}

@article{Lunine1985,
  title={Thermodynamics of clathrate hydrate at low and high pressures with application to the outer solar system},
  author={Lunine, Jonathan I and Stevenson, David J},
  journal={ApJS},
  volume={58},
  pages={493--531},
  year={1985}
}

@ARTICLE{Mousis2009a,
       author = {{Mousis}, Olivier and {Lunine}, Jonathan I. and {Thomas}, Caroline and {Pasek}, Matthew and {Marb{\oe}uf}, Ulysse and {Alibert}, Yann and {Ballenegger}, Vincent and {Cordier}, Daniel and {Ellinger}, Yves and {Pauzat}, Fran{\c{c}}oise and {Picaud}, Sylvain},
        title = "{Clathration of Volatiles in the Solar Nebula and Implications for the Origin of Titan's Atmosphere}",
      journal = {ApJ},
         year = 2009,
       volume = {691},
       number = {2},
        pages = {1780-1786},
archivePrefix = {arXiv},
       eprint = {0810.0308},
 primaryClass = {astro-ph},
       adsurl = {https://ui.adsabs.harvard.edu/abs/2009ApJ...691.1780M}
}

@ARTICLE{Leroy2015,
       author = {{Le Roy}, L{\'e}na and {Altwegg}, Kathrin and {Balsiger}, Hans and {Berthelier}, Jean-Jacques and {Bieler}, Andre and {Briois}, Christelle and {Calmonte}, Ursina and {Combi}, Michael R. and {De Keyser}, Johan and {Dhooghe}, Frederik and {Fiethe}, Bj{\"o}rn and {Fuselier}, Stephen A. and {Gasc}, S{\'e}bastien and {Gombosi}, Tamas I. and {H{\"a}ssig}, Myrtha and {J{\"a}ckel}, Annette and {Rubin}, Martin and {Tzou}, Chia-Yu},
        title = "{Inventory of the volatiles on comet 67P/Churyumov-Gerasimenko from Rosetta/ROSINA}",
      journal = {A\&A},
         year = 2015,
       volume = {583},
          eid = {A1},
        pages = {A1},
       adsurl = {https://ui.adsabs.harvard.edu/abs/2015A&A...583A...1L}
}

@ARTICLE{Mousis2014,
       author = {{Mousis}, Olivier and {Lunine}, Jonathan I. and {Fletcher}, Leigh N. and {Mandt}, Kathleen E. and {Ali-Dib}, Mohamad and {Gautier}, Daniel and {Atreya}, Sushil},
        title = "{New Insights on Saturn's Formation from its Nitrogen Isotopic Composition}",
      journal = {ApJL},
         year = 2014,
       volume = {796},
       number = {2},
          eid = {L28},
        pages = {L28},
archivePrefix = {arXiv},
       eprint = {1410.5408},
 primaryClass = {astro-ph.EP},
       adsurl = {https://ui.adsabs.harvard.edu/abs/2014ApJ...796L..28M}
}

@ARTICLE{Paardekooper2018,
       author = {{Paardekooper}, Sijme-Jan and {Johansen}, Anders},
        title = "{Giant Planet Formation and Migration}",
      journal = {Space Sci. Rev.},
         year = 2018,
       volume = {214},
       number = {1},
          eid = {38},
        pages = {38},
       adsurl = {https://ui.adsabs.harvard.edu/abs/2018SSRv..214...38P}
}

@ARTICLE{Helled2014,
       author = {{Helled}, Ravit and {Bodenheimer}, Peter},
        title = "{The Formation of Uranus and Neptune: Challenges and Implications for Intermediate-mass Exoplanets}",
      journal = {ApJ},
         year = 2014,
       volume = {789},
       number = {1},
          eid = {69},
        pages = {69},
archivePrefix = {arXiv},
       eprint = {1404.5018},
 primaryClass = {astro-ph.EP},
       adsurl = {https://ui.adsabs.harvard.edu/abs/2014ApJ...789...69H}
}

@article{Agundez2014,
	adsurl = {http://cdsads.u-strasbg.fr/abs/2014A%26A...564A..73A},
	archiveprefix = {arXiv},
	author = {{Ag{\'u}ndez}, M. and {Parmentier}, V. and {Venot}, O. and {Hersant}, F. and {Selsis}, F.},
	eid = {A73},
	eprint = {1403.0121},
	journal = {A\&A},
	pages = {A73},
	primaryclass = {astro-ph.EP},
	title = {{Pseudo 2D chemical model of hot-Jupiter atmospheres: application to HD 209458b and HD 189733b}},
	volume = 564,
	year = 2014}

@article{agundez2026,
  title={Quantification of abundance uncertainties in chemical models of exoplanet atmospheres},
  author={Ag{\'u}ndez, Marcelino},
  journal={A\&A},
  volume={705},
  pages={A50},
  year={2026},
  publisher={EDP Sciences}
}

@article{archinal2018,
  title={Report of the IAU working group on cartographic coordinates and rotational elements: 2015},
  author={Archinal, BA and Acton, CH and A’hearn, MF and Conrad, A and Consolmagno, GJ and Duxbury, T and Hestroffer, D and Hilton, JL and Kirk, Randolph L and Klioner, SA and others},
  journal={Celest. Mech. Dyn. Astron.},
  volume={130},
  number={3},
  pages={22},
  year={2018},
  publisher={Springer}
}

@article{Atreya2020,
  title={Deep atmosphere composition, structure, origin, and exploration, with particular focus on critical in situ science at the icy giants},
  author={Atreya, Sushil K and Hofstadter, Mark H and In, Joong Hyun and Mousis, Olivier and Reh, Kim and Wong, Michael H},
  journal={Space Sci. Rev.},
  volume={216},
  pages={1--31},
  year={2020},
  publisher={Springer}
}

@article{Boss2002,
  title={Rapid formation of ice giant planets},
  author={Boss, Alan P and Wetherill, George W and Haghighipour, Nader},
  journal={Icarus},
  volume={156},
  number={1},
  pages={291--295},
  year={2002},
  publisher={Elsevier}
}

@article{Burgdorf2003,
  title={Neptune's far-infrared spectrum from the ISO long-wavelength and short-wavelength spectrometers},
  author={Burgdorf, Martin and Orton, Glenn S and Davis, Gary R and Sidher, Sunil D and Feuchtgruber, Helmut and Griffin, Matthew J and Swinyard, Bruce M},
  journal={Icarus},
  volume={164},
  number={1},
  pages={244--253},
  year={2003},
  publisher={Elsevier}
}

@article{Cavalie2014,
	adsurl = {http://adsabs.harvard.edu/abs/2014A%26A...562A..33C},
	archiveprefix = {arXiv},
	author = {{Cavali{\'e}}, T. and {Moreno}, R. and {Lellouch}, E. and {Hartogh}, P. and {Venot}, O. and {Orton}, G.~S. and {Jarchow}, C. and {Encrenaz}, T. and {Selsis}, F. and {Hersant}, F. and {Fletcher}, L.~N.},
	eid = {A33},
	eprint = {1311.2458},
	journal = {A\&A},
	pages = {A33},
	primaryclass = {astro-ph.EP},
	title = {{The first submillimeter observation of CO in the stratosphere of Uranus}},
	volume = 562,
	year = 2014}

@article{Cavalie2017,
  title={Thermochemistry and vertical mixing in the tropospheres of Uranus and Neptune: How convection inhibition can affect the derivation of deep oxygen abundances},
  author={Cavali{\'e}, Thibault and Venot, Olivia and Selsis, Franck and Hersant, Franck and Hartogh, Paul and Leconte, J{\'e}r{\'e}my},
  journal={Icarus},
  volume={291},
  pages={1--16},
  year={2017},
  publisher={Elsevier}
}

@article{Cavalie2020,
	adsurl = {https://ui.adsabs.harvard.edu/abs/2020SSRv..216...58C},
	archiveprefix = {arXiv},
	author = {{Cavali{\'e}}, Thibault and {Venot}, Olivia and {Miguel}, Yamila and {Fletcher}, Leigh N. and {Wurz}, Peter and {Mousis}, Olivier and {Bounaceur}, Roda and {Hue}, Vincent and {Leconte}, J{\'e}r{\'e}my and {Dobrijevic}, Michel},
	eid = {58},
	eprint = {2004.13987},
	journal = {Space Sci. Rev.},
	number = {4},
	pages = {58},
	primaryclass = {astro-ph.EP},
	title = {{The Deep Composition of Uranus and Neptune from In Situ Exploration and Thermochemical Modeling}},
	volume = {216},
	year = 2020}

@ARTICLE{Cavalie2023,
       author = {{Cavali{\'e}}, T. and {Lunine}, J. and {Mousis}, O.},
        title = "{A subsolar oxygen abundance or a radiative region deep in Jupiter revealed by thermochemical modelling}",
      journal = {Nat. Astron.},
         year = 2023,
       volume = {7},
        pages = {678-683},
archivePrefix = {arXiv},
       eprint = {2305.13949},
 primaryClass = {astro-ph.EP},
       adsurl = {https://ui.adsabs.harvard.edu/abs/2023NatAs...7..678C}
}

@article{clement2024,
  title={Storms and convection on Uranus and Neptune: impact of methane abundance revealed by a 3D cloud-resolving model},
  author={Cl{\'e}ment, No{\'e} and Leconte, J{\'e}r{\'e}my and Spiga, Aymeric and Guerlet, Sandrine and Selsis, Franck and Milcareck, Gwena{\"e}l and Teinturier, Lucas and Cavali{\'e}, Thibault and Moreno, Rapha{\"e}l and Lellouch, Emmanuel and others},
  journal={A\&A},
  volume={690},
  pages={A227},
  year={2024},
  publisher={EDP Sciences}
}

@article{Conrath1987,
	adsurl = {http://cdsads.u-strasbg.fr/abs/1987JGR....9215003C},
	author = {{Conrath}, B. and {Hanel}, R. and {Gautier}, D. and {Marten}, A. and {Lindal}, G.},
	journal = {J. Geophys. Res.},
	pages = {15003-15010},
	title = {{The helium abundance of Uranus from Voyager measurements}},
	volume = 92,
	year = 1987}

@article{Conrath2000,
	adsurl = {http://cdsads.u-strasbg.fr/abs/2000Icar..144..124C},
	author = {{Conrath}, B.~J. and {Gautier}, D.},
	journal = {Icarus},
	pages = {124-134},
	title = {{Saturn Helium Abundance: A Reanalysis of Voyager Measurements}},
	volume = 144,
	year = 2000}

@article{deleuil2020,
  title={Observational constraints on the formation and evolution of Neptune-class exoplanets},
  author={Deleuil, Magali and Pollacco, Don and Baruteau, Cl{\'e}ment and Rauer, Heike and Blanc, Michel},
  journal={Space Sci. Rev.},
  volume={216},
  number={6},
  pages={105},
  year={2020},
  publisher={Springer}
}

@article{desch1986,
  title={The rotation period of Uranus},
  author={Desch, MD and Connerney, JEP and Kaiser, ML},
  journal={Nature},
  volume={322},
  number={6074},
  pages={42--43},
  year={1986},
  publisher={Nature Publishing Group UK London}
}

@ARTICLE{Benne2022,
       author = {{Benne}, B. and {Dobrijevic}, M. and {Cavali{\'e}}, T. and {Loison}, J.-C. and {Hickson}, K.~M.},
        title = "{A photochemical model of Triton's atmosphere paired with an uncertainty propagation study}",
      journal = {A\&A},
         year = 2022,
       volume = {667},
          eid = {A169},
        pages = {A169},
archivePrefix = {arXiv},
       eprint = {2209.04324},
 primaryClass = {astro-ph.EP},
       adsurl = {https://ui.adsabs.harvard.edu/abs/2022A&A...667A.169B}
}

@article{Dobrijevic1998,
	adsurl = {http://cdsads.u-strasbg.fr/abs/1998P%26SS...46..491D},
	author = {{Dobrijevic}, M. and {Parisot}, J.~P.},
	journal = {Planet. Space Sci.},
	pages = {491-505},
	title = {{Effect of chemical kinetics uncertainties on hydrocarbon production in the stratosphere of Neptune}},
	volume = 46,
	year = 1998}

@article{dobrijevic2010neptune,
	adsurl = {http://cdsads.u-strasbg.fr/abs/2010P%26SS...58.1555D},
	author = {{Dobrijevic}, M. and {Cavali{\'e}}, T. and {H{\'e}brard}, E. and {Billebaud}, F. and {Hersant}, F. and {Selsis}, F.},
	journal = {Planet. Space Sci.},
	pages = {1555-1566},
	title = {{Key reactions in the photochemistry of hydrocarbons in Neptune's stratosphere}},
	volume = 58,
	year = 2010}

@article{Dobrijevic2011,
	adsurl = {http://cdsads.u-strasbg.fr/abs/2011Icar..214..275D},
	author = {{Dobrijevic}, M. and {Cavali{\'e}}, T. and {Billebaud}, F.},
	journal = {Icarus},
	pages = {275-285},
	title = {{A methodology to construct a reduced chemical scheme for 2D-3D photochemical models: Application to Saturn}},
	volume = 214,
	year = 2011}

@article{encrenaz1996b,
  title={Millimeter Spectroscopy of Uranus and Neptune: Constraints on CO and PH3Tropospheric Abundances},
  author={Encrenaz, Th and Serabyn, E and Weisstein, EW},
  journal={Icarus},
  volume={124},
  number={2},
  pages={616--624},
  year={1996},
  publisher={Elsevier}
}

@article{Fernando1991,
  title={Effects of rotation on convective turbulence},
  author={Fernando, Harindra JS and Chen, Rui-Rong and Boyer, Don L},
  journal={J. Fluid Mech.},
  volume={228},
  pages={513--547},
  year={1991},
  publisher={Cambridge University Press}
}

@article{Fletcher2009b,
	author = {L.N. Fletcher and G.S. Orton and P. Yanamandra-Fisher and B.M. Fisher and P.D. Parrish and P.G.J. Irwin},
	journal = {Icarus},
	number = {1},
	pages = {154 - 175},
	title = {Retrievals of atmospheric variables on the gas giants from ground-based mid-infrared imaging},
	volume = {200},
	year = {2009}}

@article{Freedman2008,
  title={Line and mean opacities for ultracool dwarfs and extrasolar planets},
  author={Freedman, Richard S and Marley, Mark S and Lodders, Katharina},
  journal={ApJS},
  volume={174},
  number={2},
  pages={504},
  year={2008},
  publisher={IOP Publishing}
}

@article{Gao2016,
  title={Reaction Mechanism Generator: Automatic construction of chemical kinetic mechanisms},
  author={Gao, Connie W and Allen, Joshua W and Green, William H and West, Richard H},
  journal = {Comput. Phys. Commun.},
  volume={203},
  pages={212--225},
  year={2016},
  publisher={Elsevier}
}

@article{Ge2024,
  title={Heat-flux-limited cloud activity and vertical mixing in giant planet atmospheres with an application to Uranus and Neptune},
  author={Ge, Huazhi and Li, Cheng and Zhang, Xi and Moeckel, Chris},
  journal={Planet. Sci. J.},
  volume={5},
  number={4},
  pages={101},
  year={2024},
  publisher={IOP Publishing}
}

@article{ge2025,
  title={Nonuniform water distribution in Jupiter’s midlatitudes: Influence of precipitation and planetary rotation},
  author={Ge, Huazhi and Li, Cheng and Zhang, Xi and Ingersoll, Andrew P and Chen, Sihe},
  journal={Proc. Natl. Acad. Sci.},
  volume={122},
  number={41},
  pages={e2419087122},
  year={2025},
  publisher={National Academy of Sciences}
}

@phdthesis{ge2023moist,
  title={Moist Convection and Weather on Giant Planets},
  author={Ge, Huazhi},
  year={2023},
  school={University of California, Santa Cruz}
}

@article{grassi2020,
  title={On the spatial distribution of minor species in Jupiter's troposphere as inferred from Juno JIRAM data},
  author={Grassi, Davide and Adriani, Alberto and Mura, Alessandro and Atreya, SK and Fletcher, LN and Lunine, JI and Orton, GS and Bolton, S and Plainaki, CHRISTINA and Sindoni, Giuseppe and others},
  journal={J. Geophys. Res.},
  volume={125},
  number={4},
  pages={e2019JE006206},
  year={2020},
  publisher={Wiley Online Library}
}

@article{Guerlet2014,
	adsurl = {http://cdsads.u-strasbg.fr/abs/2014Icar..238..110G},
	author = {{Guerlet}, S. and {Spiga}, A. and {Sylvestre}, M. and {Indurain}, M. and {Fouchet}, T. and {Leconte}, J. and {Millour}, E. and {Wordsworth}, R. and {Capderou}, M. and {B{\'e}zard}, B. and {Forget}, F.},
	journal = {Icarus},
	pages = {110-124},
	title = {{Global climate modeling of Saturn's atmosphere. Part I: Evaluation of the radiative transfer model}},
	volume = 238,
	year = 2014}

@article{Guillot1995,
  title={Condensation of methane, ammonia, and water and the inhibition of convection in giant planets},
  author={Guillot, Tristan},
  journal={Science},
  volume={269},
  number={5231},
  pages={1697--1699},
  year={1995},
  publisher={American Association for the Advancement of Science}
}

@article{Guillot2005,
  title={The interiors of giant planets: Models and outstanding questions},
  author={Guillot, Tristan},
  journal={Annu. Rev. Earth Planet. Sci.},
  volume={33},
  number={1},
  pages={493--530},
  year={2005},
  publisher={Annual Reviews}
}

@article{helled2020uranus,
  title={Uranus and Neptune: origin, evolution and internal structure},
  author={Helled, Ravit and Nettelmann, Nadine and Guillot, Tristan},
  journal={Space Sci. Rev.},
  volume={216},
  number={3},
  pages={38},
  year={2020},
  publisher={Springer}
}

@article{Hidaka1989,
  title={Thermal decomposition of methanol in shock waves},
  author={Hidaka, Yoshiaki and Oki, Takashi and Kawano, Hiroyuki and Higashihara, Tetsuo},
  journal={J. Phys. Chem.},
  volume={93},
  number={20},
  pages={7134--7139},
  year={1989},
  publisher={ACS Publications}
}

@article{Hue2015,
	adsurl = {https://ui.adsabs.harvard.edu/abs/2015Icar..257..163H},
	archiveprefix = {arXiv},
	author = {{Hue}, V. and {Cavali{\'e}}, T. and {Dobrijevic}, M. and {Hersant}, F. and {Greathouse}, T.~K.},
	eprint = {1504.02326},
	journal = {Icarus},
	pages = {163-184},
	primaryclass = {astro-ph.EP},
	title = {{2D photochemical modeling of Saturn's stratosphere. Part I: Seasonal variation of atmospheric composition without meridional transport}},
	volume = {257},
	year = 2015}

@article{Hue2018,
	adsurl = {https://ui.adsabs.harvard.edu/abs/2018Icar..307..106H},
	archiveprefix = {arXiv},
	author = {{Hue}, V. and {Hersant}, F. and {Cavali{\'e}}, T. and {Dobrijevic}, M. and {Sinclair}, J.~A.},
	eprint = {1802.08697},
	journal = {Icarus},
	pages = {106-123},
	primaryclass = {astro-ph.EP},
	title = {{Photochemistry, mixing and transport in Jupiter's stratosphere constrained by Cassini}},
	volume = {307},
	year = 2018}

@article{Hueso2019,
  title={Atmospheric dynamics and vertical structure of Uranus and Neptune’s weather layers},
  author={Hueso, Ricardo and S{\'a}nchez-Lavega, Agust{\'\i}n},
  journal={Space Sci. Rev.},
  volume={215},
  number={8},
  pages={52},
  year={2019},
  publisher={Springer}
}

@article{hyder2025supersolar,
  title={A supersolar oxygen abundance supported by hydrodynamic modelling of Jupiter’s atmosphere},
  author={Hyder, Ali and Li, Cheng and Chanover, Nancy and Bjoraker, Gordon},
  journal={Nat. Astron.},
  volume={9},
  number={2},
  pages={211--220},
  year={2025},
  publisher={Nature Publishing Group UK London}
}

@article{Irwin2018,
  title={Detection of hydrogen sulfide above the clouds in Uranus’s atmosphere},
  author={Irwin, Patrick GJ and Toledo, Daniel and Garland, Ryan and Teanby, Nicholas A and Fletcher, Leigh N and Orton, Glenn A and B{\'e}zard, Bruno},
  journal={Nat. Astron.},
  volume={2},
  number={5},
  pages={420--427},
  year={2018},
  publisher={Nature Publishing Group UK London}
}

@article{Irwin2019,
  title={Probable detection of hydrogen sulphide (H2S) in Neptune’s atmosphere},
  author={Irwin, Patrick GJ and Toledo, Daniel and Garland, Ryan and Teanby, Nicholas A and Fletcher, Leigh N and Orton, Glenn S and B{\'e}zard, Bruno},
  journal={Icarus},
  volume={321},
  pages={550--563},
  year={2019},
  publisher={Elsevier}
}

@article{Irwin2021,
  title={Latitudinal variation of methane mole fraction above clouds in Neptune's atmosphere from VLT/MUSE-NFM: Limb-darkening reanalysis},
  author={Irwin, Patrick GJ and Dobinson, Jack and James, Arjuna and Toledo, Daniel and Teanby, Nicholas A and Fletcher, Leigh N and Orton, Glenn S and P{\'e}rez-Hoyos, Santiago},
  journal={Icarus},
  volume={357},
  pages={114277},
  year={2021},
  publisher={Elsevier}
}

@article{Irwin2022,
  title={Hazy blue worlds: a holistic aerosol model for Uranus and Neptune, including dark spots},
  author={Irwin, Patrick GJ and Teanby, Nicholas A and Fletcher, Leigh N and Toledo, Daniel and Orton, Glenn S and Wong, Michael H and Roman, Michael T and P{\'e}rez-Hoyos, Santiago and James, Arjuna and Dobinson, Jack},
  journal={J. Geophys. Res.},
  volume={127},
  number={6},
  pages={e2022JE007189},
  year={2022},
  publisher={Wiley Online Library}
}

@article{jacobson2009,
  title={The orbits of the Neptunian satellites and the orientation of the pole of Neptune},
  author={Jacobson, Robert A},
  journal={AJ},
  volume={137},
  number={5},
  pages={4322},
  year={2009},
  publisher={IOP Publishing}
}

@article{jacobson2014,
  title={The orbits of the Uranian satellites and rings, the gravity field of the Uranian system, and the orientation of the pole of Uranus},
  author={Jacobson, RA},
  journal={AJ},
  volume={148},
  number={5},
  pages={76},
  year={2014},
  publisher={IOP Publishing}
}

@article{lecacheux1993,
  title={The sidereal rotation period of Neptune},
  author={Lecacheux, Alain and Zarka, Ph and Desch, MD and Evans, DR},
  journal={Geophys. Res. Lett.},
  volume={20},
  number={23},
  pages={2711--2714},
  year={1993},
  publisher={Wiley Online Library}
}

@article{Leconte2017,
  title={Condensation-inhibited convection in hydrogen-rich atmospheres-Stability against double-diffusive processes and thermal profiles for Jupiter, Saturn, Uranus, and Neptune},
  author={Leconte, J{\'e}r{\'e}my and Selsis, Franck and Hersant, Franck and Guillot, Tristan},
  journal={A\&A},
  volume={598},
  pages={A98},
  year={2017},
  publisher={EDP Sciences}
}

@article{Lellouch2005,
	adsurl = {http://adsabs.harvard.edu/abs/2005A%26A...430L..37L},
	author = {{Lellouch}, E. and {Moreno}, R. and {Paubert}, G.},
	journal = {A\&A},
	pages = {L37-L40},
	title = {{A dual origin for Neptune's carbon monoxide?}},
	volume = 430,
	year = 2005}

@article{Li2020,
	author = {Li, W. and Shen, X. -C. and Menietti, J. D. and Ma, Q. and Zhang, X. -J. and Kurth, W. S. and Hospodarsky, G. B.},
	c7 = {e2020GL088198},
	c8 = {2020GL088198},
	isbn = {0094-8276},
	journal = {Geophys. Res. Lett.},
	journal1 = {Geophys. Res. Lett.},
	number = {15},
	pages = {e2020GL088198},
	title = {Global Distribution of Whistler Mode Waves in Jovian Inner Magnetosphere},
	ty = {JOUR},
	volume = {47},
	year = {2020}}

@article{Lindal1992,
  title={The atmosphere of Neptune-an analysis of radio occultation data acquired with Voyager 2},
  author={Lindal, Gunnar F},
  journal={AJ},
  volume={103},
  pages={967--982},
  year={1992}
}

@article{lodders2025solar,
  title={Solar system elemental abundances from the solar photosphere and CI-chondrites},
  author={Lodders, Katharina and Bergemann, Maria and Palme, Herbert},
  journal={Space Sci. Rev.},
  volume={221},
  number={2},
  pages={23},
  year={2025},
  publisher={Springer}
}

@article{Luszcz2013,
  title={Constraining the origins of Neptune’s carbon monoxide abundance with CARMA millimeter-wave observations},
  author={Luszcz-Cook, Statia H and de Pater, Imke},
  journal={Icarus},
  volume={222},
  number={1},
  pages={379--400},
  year={2013},
  publisher={Elsevier}
}

@article{milcareck2024,
  title={Radiative-convective models of the atmospheres of Uranus and Neptune: heating sources and seasonal effects},
  author={Milcareck, Gwena{\"e}l and Guerlet, Sandrine and Montmessin, Franck and Spiga, Aymeric and Leconte, Jeremy and Millour, Ehouarn and Clement, Noe and Fletcher, Leigh N and Roman, Michael T and Lellouch, Emmanuel and others},
  journal={A\&A},
  volume={686},
  pages={A303},
  year={2024},
  publisher={EDP Sciences}
}

@article{Molter2021,
  title={Tropospheric composition and circulation of Uranus with ALMA and the VLA},
  author={Molter, Edward M and De Pater, Imke and Luszcz-Cook, Statia and Tollefson, Joshua and Sault, Robert J and Butler, Bryan and De Boer, David},
  journal={Planet. Sci. J.},
  volume={2},
  number={1},
  pages={3},
  year={2021},
  publisher={IOP Publishing}
}

@article{Moreno2017,
	adsurl = {http://cdsads.u-strasbg.fr/abs/2017A%26A...608L...5M},
	author = {{Moreno}, R. and {Lellouch}, E. and {Cavali{\'e}}, T. and {Moullet}, A.},
	eid = {L5},
	journal = {A\&A},
	pages = {L5},
	title = {{Detection of CS in Neptune's atmosphere from ALMA observations}},
	volume = 608,
	year = 2017}

@article{Mousis2010,
  title={Volatile inventories in clathrate hydrates formed in the primordial nebula},
  author={Mousis, Olivier and Lunine, Jonathan I and Picaud, Sylvain and Cordier, Daniel},
  journal={Faraday Discuss.},
  volume={147},
  pages={509--525},
  year={2010},
  publisher={Royal Society of Chemistry}
}

@article{Mousis2020,
	adsurl = {https://ui.adsabs.harvard.edu/abs/2020RSPTA.37800107M},
	author = {{Mousis}, O. and {Aguichine}, A. and {Helled}, R. and {Irwin}, P.~G.~J. and {Lunine}, J.~I.},
	eid = {20200107},
	journal = {Philos. Trans. R. Soc. A},
	number = {2187},
	pages = {20200107},
	title = {{The role of ice lines in the formation of Uranus and Neptune}},
	volume = {378},
	year = 2020}

@article{Schneeberger2023,
	adsurl = {https://ui.adsabs.harvard.edu/abs/2023A&A...670A..28S},
	archiveprefix = {arXiv},
	author = {{Schneeberger}, Antoine and {Mousis}, Olivier and {Aguichine}, Artyom and {Lunine}, Jonathan I.},
	eid = {A28},
	eprint = {2301.02482},
	journal = {A\&A},
	pages = {A28},
	primaryclass = {astro-ph.EP},
	title = {{Evolution of the reservoirs of volatiles in the protosolar nebula}},
	volume = {670},
	year = 2023}

@article{Moses2020,
  title={Atmospheric chemistry on Uranus and Neptune},
  author={Moses, Julianne I and Cavali{\'e}, T and Fletcher, LN and Roman, MT},
  journal={Philos. Trans. R. Soc. A},
  volume={378},
  number={2187},
  pages={20190477},
  year={2020},
  publisher={The Royal Society Publishing}
}

@article{Mousis2018,
	adsurl = {https://ui.adsabs.harvard.edu/abs/2018P&SS..155...12M},
	archiveprefix = {arXiv},
	author = {{Mousis}, O. and {Atkinson}, D.~H. and {Cavali{\'e}}, T. and {Fletcher}, L.~N. and {Amato}, M.~J. and {Aslam}, S. and {Ferri}, F. and {Renard}, J. -B. and {Spilker}, T. and {Venkatapathy}, E. and {Wurz}, P. and {Aplin}, K. and {Coustenis}, A. and {Deleuil}, M. and {Dobrijevic}, M. and {Fouchet}, T. and {Guillot}, T. and {Hartogh}, P. and {Hewagama}, T. and {Hofstadter}, M.~D. and {Hue}, V. and {Hueso}, R. and {Lebreton}, J. -P. and {Lellouch}, E. and {Moses}, J. and {Orton}, G.~S. and {Pearl}, J.~C. and {S{\'a}nchez-Lavega}, A. and {Simon}, A. and {Venot}, O. and {Waite}, J.~H. and {Achterberg}, R.~K. and {Atreya}, S. and {Billebaud}, F. and {Blanc}, M. and {Borget}, F. and {Brugger}, B. and {Charnoz}, S. and {Chiavassa}, T. and {Cottini}, V. and {d'Hendecourt}, L. and {Danger}, G. and {Encrenaz}, T. and {Gorius}, N.~J.~P. and {Jorda}, L. and {Marty}, B. and {Moreno}, R. and {Morse}, A. and {Nixon}, C. and {Reh}, K. and {Ronnet}, T. and {Schmider}, F. -X. and {Sheridan}, S. and {Sotin}, C. and {Vernazza}, P. and {Villanueva}, G.~L.},
	eprint = {1708.00235},
	journal = {Planet. Space Sci.},
	pages = {12-40},
	primaryclass = {astro-ph.EP},
	title = {{Scientific rationale for Uranus and Neptune in situ explorations}},
	volume = {155},
	year = 2018}

@article{Mousis2024,
  title={Insights on the Formation Conditions of Uranus and Neptune from Their Deep Elemental Compositions},
  author={Mousis, Olivier and Schneeberger, Antoine and Cavali{\'e}, Thibault and Mandt, Kathleen E and Aguichine, Artyom and Lunine, Jonathan I and Benest Couzinou, Tom and Hue, Vincent and Moreno, Rapha{\"e}l},
  journal={Planet. Sci. J.},
  volume={5},
  number={8},
  pages={173},
  year={2024},
  publisher={The American Astronomical Society}
}

@article{Birnstiel2012,
	adsurl = {https://ui.adsabs.harvard.edu/abs/2012A&A...539A.148B},
	archiveprefix = {arXiv},
	author = {{Birnstiel}, T. and {Klahr}, H. and {Ercolano}, B.},
	eid = {A148},
	eprint = {1201.5781},
	journal = {A\&A},
	pages = {A148},
	primaryclass = {astro-ph.EP},
	title = {{A simple model for the evolution of the dust population in protoplanetary disks}},
	volume = {539},
	year = 2012}

@article{Niemann1998,
	adsurl = {http://cdsads.u-strasbg.fr/abs/1998JGR...10322831N},
	author = {{Niemann}, H.~B. and {Atreya}, S.~K. and {Carignan}, G.~R. and {Donahue}, T.~M. and {Haberman}, J.~A. and {Harpold}, D.~N. and {Hartle}, R.~E. and {Hunten}, D.~M. and {Kasprzak}, W.~T. and {Mahaffy}, P.~R. and {Owen}, T.~C. and {Way}, S.~H.},
	journal = {J. Geophys. Res.},
	pages = {22831-22846},
	title = {{The composition of the Jovian atmosphere as determined by the Galileo probe mass spectrometer}},
	volume = 103,
	year = 1998}

@article{Orton2014,
  title={Mid-infrared spectroscopy of Uranus from the Spitzer Infrared Spectrometer: 1. Determination of the mean temperature structure of the upper troposphere and stratosphere},
  author={Orton, Glenn S and Fletcher, Leigh N and Moses, Julianne I and Mainzer, Amy K and Hines, Dean and Hammel, Heidi B and Martin-Torres, F Javier and Burgdorf, Martin and Merlet, Cecile and Line, Michael R},
  journal={Icarus},
  volume={243},
  pages={494--513},
  year={2014},
  publisher={Elsevier}
}

@article{Owen1999,
  title={A low-temperature origin for the planetesimals that formed Jupiter},
  author={Owen, Tobias and Mahaffy, Paul and Niemann, HB and Atreya, Sushil and Donahue, Thomas and Bar-Nun, Akiva and de Pater, Imke},
  journal={Nature},
  volume={402},
  number={6759},
  pages={269--270},
  year={1999},
  publisher={Nature Publishing Group UK London}
}

@article{pearl1991albedo,
  title={The albedo, effective temperature, and energy balance of Neptune, as determined from Voyager data},
  author={Pearl, JC and Conrath, BJ},
  journal={J. Geophys. Res.},
  volume={96},
  number={S01},
  pages={18921--18930},
  year={1991},
  publisher={Wiley Online Library}
}

@article{Prinn1977,
  title={Carbon monoxide on Jupiter and implications for atmospheric convection},
  author={Prinn, Ronald G and Barshay, Stephen S},
  journal={Science},
  volume={198},
  number={4321},
  pages={1031--1034},
  year={1977},
  publisher={American Association for the Advancement of Science}
}

@article{Prinn1981,
  title={Kinetic inhibition of CO and N2 reduction in circumplanetary nebulae-Implications for satellite composition},
  author={Prinn, RONALD G and Fegley Jr, Bruce},
  journal={ApJ},
  volume={249},
  pages={308--317},
  year={1981}
}

@INPROCEEDINGS{Raymond2022,
       author = {{Raymond}, Sean N. and {Morbidelli}, Alessandro},
        title = "{Planet Formation: Key Mechanisms and Global Models}",
    booktitle = {Demographics of Exoplanetary Systems, Lecture Notes of the 3rd Advanced School on Exoplanetary Science},
         year = 2022,
       editor = {{Biazzo}, Katia and {Bozza}, Valerio and {Mancini}, Luigi and {Sozzetti}, Alessandro},
       series = {Astrophysics and Space Science Library},
       volume = {466},
        pages = {3-82},
archivePrefix = {arXiv},
       eprint = {2002.05756},
 primaryClass = {astro-ph.EP},
       adsurl = {https://ui.adsabs.harvard.edu/abs/2022ASSL..466....3R}
}

@techreport{roman2025temperature,
  title={Temperature, Composition, and Cloud structure in Atmosphere of Uranus from MIRI-MRS and NIRSpec-IFU Spectra},
  author={Roman, Michael and Fletcher, Leigh and Hammel, Heidi and Irwin, Patrick and King, Oliver and Rowe-Gurney, Naomi and Moses, Julianne and Orton, Glenn and de Pater, Imke and Melin, Henrik and others},
  year={2025},
  institution={Copernicus Meetings}
}

@article{siebenaler2025b,
  title={Mean opacity tables for probing the interior and atmosphere of giant planets},
  author={Siebenaler, Louis and Miguel, Yamila},
  journal={MNRAS},
  pages={staf2205},
  year={2025},
  publisher={Oxford University Press}
}

@article{Smith1998,
	adsurl = {https://ui.adsabs.harvard.edu/abs/1998Icar..132..176S},
	author = {{Smith}, Michael D.},
	journal = {Icarus},
	number = {1},
	pages = {176-184},
	title = {{Estimation of a Length Scale to Use with the Quench Level Approximation for Obtaining Chemical Abundances}},
	volume = {132},
	year = 1998}

@article{Sromovsky2014,
	adsurl = {http://cdsads.u-strasbg.fr/abs/2014Icar..238..137S},
	archiveprefix = {arXiv},
	author = {{Sromovsky}, L.~A. and {Karkoschka}, E. and {Fry}, P.~M. and {Hammel}, H.~B. and {de Pater}, I. and {Rages}, K.},
	eprint = {1502.06480},
	journal = {Icarus},
	pages = {137-155},
	primaryclass = {astro-ph.EP},
	title = {{Methane depletion in both polar regions of Uranus inferred from HST/STIS and Keck/NIRC2 observations}},
	volume = 238,
	year = 2014}

@article{Sromovsky2019,
  title={The methane distribution and polar brightening on Uranus based on HST/STIS, Keck/NIRC2, and IRTF/SpeX observations through 2015},
  author={Sromovsky, Lawrence A and Karkoschka, Erich and Fry, Patrick M and de Pater, Imke and Hammel, Heidi B},
  journal={Icarus},
  volume={317},
  pages={266--306},
  year={2019},
  publisher={Elsevier}
}

@inproceedings{Stone1976,
  title={The meteorology of the Jovian atmosphere},
  author={Stone, PH},
  booktitle={IAU Colloq. 30: Jupiter: Studies of the Interior, Atmosp here, Magnetosphere and Satellites},
  pages={586--618},
  year={1976}
}

@article{Teanby2013,
  title={An external origin for carbon monoxide on Uranus from Herschel/SPIRE?},
  author={Teanby, NA and Irwin, PGJ},
  journal={ApJL},
  volume={775},
  number={2},
  pages={L49},
  year={2013},
  publisher={IOP Publishing}
}

@article{Teanby2019,
  title={Neptune’s carbon monoxide profile and phosphine upper limits from Herschel/SPIRE: implications for interior structure and formation},
  author={Teanby, NA and Irwin, PGJ and Moses, JI},
  journal={Icarus},
  volume={319},
  pages={86--98},
  year={2019},
  publisher={Elsevier}
}

@article{Toledo2025,
  title={Methane precipitation in ice giant atmospheres},
  author={Toledo, D and Rannou, Pascal and Irwin, Patrick and de Trenquell{\'e}on, B de Batz and Roman, Michael and Apestigue, Victor and Arruego, Ignacio and Yela, Margarita},
  journal={A\&A},
  volume={694},
  pages={A81},
  year={2025},
  publisher={EDP Sciences}
}

@article{tollefson2018,
  title={Vertical wind shear in Neptune’s upper atmosphere explained with a modified thermal wind equation},
  author={Tollefson, Joshua and de Pater, Imke and Marcus, Philip S and Luszcz-Cook, Statia and Sromovsky, Lawrence A and Fry, Patrick M and Fletcher, Leigh N and Wong, Michael H},
  journal={Icarus},
  volume={311},
  pages={317--339},
  year={2018},
  publisher={Elsevier}
}

@article{tollefson2019,
  title={Neptune's latitudinal variations as viewed with ALMA},
  author={Tollefson, Joshua and de Pater, Imke and Luszcz-Cook, Statia and DeBoer, David},
  journal={AJ},
  volume={157},
  number={6},
  pages={251},
  year={2019},
  publisher={IOP Publishing}
}

@article{Valencia2013,
  title={Bulk composition of GJ 1214b and other sub-Neptune exoplanets},
  author={Valencia, Diana and Guillot, Tristan and Parmentier, Vivien and Freedman, Richard S},
  journal={ApJ},
  volume={775},
  number={1},
  pages={10},
  year={2013},
  publisher={IOP Publishing}
}

@article{veillet2024,
  title={An extensively validated C/H/O/N chemical network for hot exoplanet disequilibrium chemistry},
  author={Veillet, R and Venot, Olivia and Sirjean, Baptiste and Bounaceur, R and Glaude, P-A and Al-Refaie, A and H{\'e}brard, E},
  journal={A\&A},
  volume={682},
  pages={A52},
  year={2024},
  publisher={EDP Sciences}
}

@article{Veillet2026,
  title={Development of a C/H/O/N/S chemical network: Experimental benchmark, application to exoplanets, and identification of key C/S coupling pathways},
  author={Veillet, R and Venot, O and Sirjean, B and Citrangolo Destro, F and Fournet, R and Al-Refaie, A and Hebrard, Eric and Glaude, P-A and Bounaceur, R},
  journal={A\&A},
  volume={706},
  pages={A260},
  year={2026},
  publisher={EDP Sciences}
}

@article{Venot2012,
	adsurl = {http://cdsads.u-strasbg.fr/abs/2012A%26A...546A..43V},
	archiveprefix = {arXiv},
	author = {{Venot}, O. and {H{\'e}brard}, E. and {Ag{\'u}ndez}, M. and {Dobrijevic}, M. and {Selsis}, F. and {Hersant}, F. and {Iro}, N. and {Bounaceur}, R.},
	eid = {A43},
	eprint = {1208.0560},
	journal = {A\&A},
	pages = {A43},
	primaryclass = {astro-ph.EP},
	title = {{A chemical model for the atmosphere of hot Jupiters}},
	volume = 546,
	year = 2012}

@article{Venot2015,
	adsurl = {http://cdsads.u-strasbg.fr/abs/2015A%26A...577A..33V},
	archiveprefix = {arXiv},
	author = {{Venot}, O. and {H{\'e}brard}, E. and {Ag{\'u}ndez}, M. and {Decin}, L. and {Bounaceur}, R.},
	eid = {A33},
	eprint = {1502.03567},
	journal = {A\&A},
	pages = {A33},
	primaryclass = {astro-ph.EP},
	title = {{New chemical scheme for studying carbon-rich exoplanet atmospheres}},
	volume = 577,
	year = 2015}

@article{venot2019reduced,
  title={Reduced chemical scheme for modelling warm to hot hydrogen-dominated atmospheres},
  author={Venot, Olivia and Bounaceur, Roda and Dobrijevic, Michel and H{\'e}brard, Eric and Cavali{\'e}, Thibault and Tremblin, Pascal and Drummond, Benjamin and Charnay, Benjamin},
  journal={A\&A},
  volume={624},
  pages={A58},
  year={2019},
  publisher={EDP Sciences}
}

@article{Venot2020,
  title={New chemical scheme for giant planet thermochemistry-Update of the methanol chemistry and new reduced chemical scheme},
  author={Venot, Olivia and Cavali{\'e}, Thibault and Bounaceur, Roda and Tremblin, Pascal and Brouillard, Lothaire and Brahim, R Lhoussaine Ben},
  journal={A\&A},
  volume={634},
  pages={A78},
  year={2020},
  publisher={EDP Sciences}
}

@phdthesis{venot2012photochimie,
  title={Photochimie des exoplanetes chaudes: mod{\'e}lisations et exp{\'e}riences},
  author={Venot, Olivia},
  year={2012},
  school={Bordeaux 1}
}

@article{Visscher2010,
  title={The deep water abundance on Jupiter: New constraints from thermochemical kinetics and diffusion modeling},
  author={Visscher, Channon and Moses, Julianne I and Saslow, Sarah A},
  journal={Icarus},
  volume={209},
  number={2},
  pages={602--615},
  year={2010},
  publisher={Elsevier}
}

@article{VonZahn1998,
	adsurl = {http://cdsads.u-strasbg.fr/abs/1998JGR...10322815V},
	author = {{von Zahn}, U. and {Hunten}, D.~M. and {Lehmacher}, G.},
	journal = {J. Geophys. Res.},
	pages = {22815-22830},
	title = {{Helium in Jupiter's atmosphere: Results from the Galileo probe helium interferometer experiment}},
	volume = 103,
	year = 1998}

@article{Wang2015,
  title={New insights on Jupiter’s deep water abundance from disequilibrium species},
  author={Wang, Dong and Gierasch, Peter J and Lunine, Jonathan I and Mousis, Olivier},
  journal={Icarus},
  volume={250},
  pages={154--164},
  year={2015},
  publisher={Elsevier}
}

@article{wang2016,
  title={Modeling the disequilibrium species for Jupiter and Saturn: Implications for Juno and Saturn entry probe},
  author={Wang, Dong and Lunine, Jonathan I and Mousis, Olivier},
  journal={Icarus},
  volume={276},
  pages={21--38},
  year={2016},
  publisher={Elsevier}
}

@article{wang2025internal,
  title={Internal heat flux and energy imbalance of Uranus},
  author={Wang, Xinyue and Li, Liming and Roman, Michael and Zhang, Xi and Jiang, Xun and Fry, Patrick and Li, Cheng and Milcareck, Gwenael and Sanchez-Lavega, Agustin and Perez-Hoyos, Santiago and others},
  journal={Geophys. Res. Lett.},
  volume={52},
  number={14},
  pages={e2025GL115660},
  year={2025},
  publisher={Wiley Online Library}
}

@article{Yang2026,
  title={Coupled 1D Chemical Kinetic Transport and 2D Hydrodynamic Modeling Supports a Modest 1--1.5$\times$ Supersolar Oxygen Abundance in Jupiter’s Atmosphere},
  author={Yang, Jeehyun and Hyder, Ali and Hu, Renyu and Lunine, Jonathan I},
  journal={Planet. Sci. J.},
  volume={7},
  number={1},
  pages={2},
  year={2026},
  publisher={IOP Publishing}
}

\newpage

\appendix
\section{Solar System abundance data}
In Fig.~\ref{fig:SS_enrichments} we present the deep elemental abundances inferred from tropospheric measurements in the giant planets of the Solar System.  The values used are given in the following table, where elemental abundances have been expressed with the most recent protosolar data reference to date \citep{lodders2025solar}.\\

\begin{table}[htbp]
\centering
\caption{Composition of giant planets of the Solar System. }
\label{table:compo_SS}
\begin{tabular}{lcccc}
\hline\hline
 & Jupiter & Saturn & Uranus & Neptune \\
\hline
He/H & $0.81 \pm 0.05^{a}$ & $0.70 \pm 0.14^{b}$ & $0.93 \pm 0.21^{f}$ & $0.93 \pm 0.16^{c}$ \\
Ne/H & $0.07 \pm 0.02^{d}$ & -- & -- & -- \\
C/H & $2.95 \pm 0.78^{g}$ & $6.65 \pm 0.41^{e}$ & $33.21$--$56.73^{f}$ & $31.82$--$41.51^{g}$ \\
N/H & $2.09 \pm 0.12^{d}$ & $1.56 \pm 1.09^{e}$ & -- & -- \\
O/H & $2.30^{+2.04\,d}_{-1.45}$ & -- & -- & -- \\
P/H & $2.87 \pm 0.32^{h}$ & $9.75 \pm 1.13^{j}$ & -- & -- \\
S/H & $2.38 \pm 0.57^{d}$ & -- & -- & -- \\
Ar/H & $2.33 \pm 0.76^{d}$ & -- & -- & -- \\
Kr/H & $1.84 \pm 0.50^{d}$ & -- & -- & -- \\
Xe/H & $1.80 \pm 0.50^{d}$ & -- & -- & -- \\
\hline
\end{tabular}

\vspace{0.3cm}

\tablefoot{Atmospheric abundances are normalized to their protosolar values, according to \cite{lodders2025solar}.\\
$^{a}$ \textit{Galileo} in situ probe \citep{VonZahn1998,Niemann1998}\\
$^{b}$ Voyager \citep{Conrath1987,Conrath2000}\\
$^{c}$ Infrared Space Observatory (ISO) \citep{Burgdorf2003}\\
$^{d}$ \textit{Galileo} in situ probe \citep{Mahaffy2000,Wong2004}\\
$^{e}$ Cassini \citep{Fletcher2009b}\\
$^{f}$ HST, Keck, and IRTF \citep{Sromovsky2019}\\
$^{g}$ ALMA \citep{tollefson2019}\\
$^{h}$ Juno \citep{Li2020}\\
$^{j}$ Cassini \citep{Fletcher2011}\\
}
\end{table}

\end{document}